%% file: ms.tex
\documentclass[prd,twocolumn,showpacs,preprintnumbers,amsmath,amssymb,nofootinbib,superscriptaddress,floatfix]{revtex4}

\usepackage{epsfig}
\usepackage{psfrag}
\usepackage{dcolumn}
\usepackage{bm}
\usepackage{hyperref}

\usepackage{subfigure}

\usepackage[usenames]{color} 

\usepackage{ifpdf}
\ifpdf
\usepackage[pdftex]{graphicx}
\else
\usepackage{graphicx}
\fi
\ifpdf
\DeclareGraphicsExtensions{.pdf, .jpg, .tif}
\else
\DeclareGraphicsExtensions{.eps, .jpg}
\fi

\def\Mpc{\,{\rm Mpc}}

\def\fun#1#2{\lower3.6pt\vbox{\baselineskip0pt\lineskip.9pt
\ialign{$\mathsurround=0pt#1\hfil##\hfil$\crcr#2\crcr\sim\crcr}}}

\newcommand{\half}{\ensuremath{\frac{1}{2}\,}}

\newcommand{\rv}{\mathbf{r}}
\newcommand{\kv}{\mathbf{k}}

\newcommand{\thetav}{\boldsymbol{\theta}}
\newcommand{\ngbar}{\bar{n}_{g}}

\newcommand{\rhonfw}{\rho_{\text{NFW}}}

\newcommand{\cla}{C_{\text{1}}}

\newcommand{\rhobar}{\bar{\rho}}

\newcommand{\eplus}{\gamma_{1}}
\newcommand{\ecross}{\gamma_{2}}
\newcommand{\intshear}{\gamma^{I}}

\newcommand{\densshear}{\tilde{\gamma}^{I}}

\newcommand{\phik}{\phi_k}

\newcommand{\sintwophie}{\sin\left(2\phi_e\right)}
\newcommand{\costwophie}{\cos\left(2\phi_e\right)}

\newcommand{\nbar}{\bar{n}}

\newcommand{\ebareff}{\ebar_{\text{scale}}}

\newcommand{\Pgi}{P_{\densshear}}
\newcommand{\fA}{\chi}

\newcommand{\eeff}{$\ebar_{\text{scale}}\,$}

\psfrag{ebarefffactor}{\eeff}

\newcommand{\ebar}{\bar{\gamma}}

\begin{document}

\title{A halo model for intrinsic alignments of
galaxy ellipticities}

\author{Michael D. Schneider}
\email{michael.schneider@durham.ac.uk}
\affiliation{Institute for Computational Cosmology, Department of Physics, Durham University, South Road, Durham, DH1 3LE, UK.}

\author{Sarah Bridle}
\email{sarah.bridle@ucl.ac.uk}
\affiliation{Department of Physics and Astronomy, University College London, Gower Street, London, WC1E 6BT, UK.}

\date{\today}

\pacs{98.80.Es, 98.62.Gq, 98.62.Sb}

\begin{abstract}
Correlations between intrinsic ellipticities of galaxies are a potentially important systematic error when constraining dark energy properties from weak gravitational lensing (cosmic shear) surveys.  In the absence of perfectly known galaxy redshifts some modeling of the galaxy intrinsic alignments  is likely to be required to extract the lensing signal to sufficient accuracy.  We present a new model based on the placement of galaxies into dark matter halos. The central galaxy ellipticity follows the large scale potential and, in the simplest case, the satellite galaxies point at the halo center.  The two-halo term is then dominated by the linear alignment model and the one-halo term provides a motivated extension of intrinsic alignment models to small scales. We provide fitting formulae for the spatial projected source power spectra for both intrinsic-intrinsic (II) and shear-intrinsic (GI) correlations.  We illustrate the potential impact of ignoring intrinsic alignments on cosmological parameter constraints from non-tomographic surveys, finding that $\sigma_8$ could be underestimated by up to the size of the current 1-$\sigma$ error bar from cosmic shear if very small scales are included in the analysis.  Finally, we highlight areas of interest for numerical simulations of dark matter clustering and galaxy formation that can further constrain the intrinsic alignment signal.
\end{abstract}

\keywords{cosmology: theory -- cosmology: weak lensing}

\maketitle

\section{Introduction} 
\label{sec:introduction}
Gravitational lensing causes the images of distant galaxies to be distorted as the light travels through space-time that has been bent by intervening matter. Two galaxies that are physically close will therefore 
appear 
preferentially aligned with each other because their light travels through similar regions of curved space-time, so 
their images 
are both distorted in a similar direction. This ``cosmic shear'' 
seems 
to be one of the most promising
probes of the nature of the mysterious dark energy~\cite{detf,esoesa}
that apparently dominates the energy budget of our universe.

However, the lensing distortions are extremely small compared to the intrinsic ellipticities of galaxies, and can only be detected by averaging the ellipticities of many spatially localized galaxies.  If galaxies are randomly oriented in space then any intrinsic ellipticities should average to zero allowing a detection of the gravitational lensing signal.  But because we believe galaxies form inside large-scale gravitational potentials, coherent tidal effects might be expected to align the intrinsic galaxy ellipticities.

There are two physical mechanisms that have been widely considered for inducing alignments of galaxy orientations.
First, there may be alignment of angular momentum vectors of galaxies that formed in the same initial tidal field due to tidal torquing~\citep[see e.g.][]{heavensp88,schaefer08}.
Second, there may be a coherent tidal stretching of galaxy shapes due to the large scale gravitational potential~\citep{catelankb01}
or anisotropic accretion along filaments~\cite{aubert04}.
The alignment of spin axes is believed to dominate for rotationally supported spiral galaxies while the tidal stretching is believed to be the dominant effect for elliptical galaxies.

Using linear tidal torque theory, Refs.~\cite{crittenden01,crittenden02} showed that the correlation function for the galaxy ellipticity orientations is nonzero only at second order in the tidal tensor and is therefore small except at small galaxy separations.  The tidal stretching of elliptical galaxies can be quantified analytically by assuming that the intrinsic ellipticity of galaxies is proportional to the curvature of the primordial large scale potential~\citep{catelankb01}. This is often referred to as the linear alignment model~\citep{hirata04}. Since the ellipticity correlation function of tidally stretched elliptical galaxies is linear in the tidal tensor it could potentially have a significant amplitude over many tens of 
megaparsecs. 

The size of these effects has been estimated using numerical simulations, in which galaxies are effectively pasted into the simulation with directions corresponding to the orientation of individual dark matter halos, or the angular momentum vector~\citep{heavens00,croftm00,leespl08,heymans06}.  This has confirmed the result that the most significant correlations come from alignments of galaxy light with the orientation of the dark matter halo~\citep{heymans06}.

On slightly smaller scales some common trends have been observed, both in simulations and observations, regarding the alignment of sub-halos or satellite galaxies within their parent halo or cluster/group. First, satellites tend to be aligned with the radius vector of their host halo~\cite{hawley75,ciotti94,pereira05,agustsson06,faltenbacher07a,knebe08,pereira08}.  Second, the satellites are found to be preferentially located near the major axis of the halo~\cite{plionis03,faltenbacher07a,agustsson07,faltenbacher08,faltenbacher09}.  Third, the satellites are aligned with the central galaxy~\cite{faltenbacher07a,faltenbacher09}.
See Ref.~\cite{siverd09} for the most recent observations of these effects.

The alignment of physically neighboring galaxies is expected to produce a spurious increase in the cosmic shear galaxy alignment signal. This is often referred to as the II (intrinsic--intrinsic) alignment. Ref.~\cite{hirata04} pointed out that an additional contamination can occur due to the simultaneous alignment of a galaxy with a nearby matter distribution, and the gravitational lensing of a more distant galaxy by the same matter distribution. This gravitational-intrinsic (GI) alignment produces a spurious decrease in the cosmic shear signal, since the two galaxies now point in opposite directions. The closer galaxy points towards the mass clump, whereas the more distant galaxy is stretched tangentially around the clump. A small-scale contribution to the GI effect even comes from the gravitational lensing of distant galaxies by a galaxy's own elliptical halo~\citep{bridlea07}.

Observations of the II effect have been carried out using low redshift galaxy
samples by
SuperCOSMOS~\citep{brownthd02},
COMBO-17~\citep{heymans04}
and SDSS~\citep{mandelbaum06,leep07}.
The raw GI effect is harder to measure, due to the large range of redshifts involved, and potential contamination by cosmic shear itself.
However the alignment between galaxies and neighboring mass has been measured by using the distribution of galaxies themselves as a tracer of the mass distribution~\citep{mandelbaum06,hirata07}.

The methods for removing both II and GI intrinsic alignment (IA) effects as a systematic error in cosmic shear measurements can be classified in two categories of ``nulling'' or ``modeling''~\cite{kitching09}.  ``Nulling'' methods downweight  selected parts of the data in such a way as to remove the intrinsic alignment signal.  For II correlations this can be done by removing galaxy pairs that are close in redshift (and on the sky)~\cite{king02,heymans03,takada04,heymans04}, while the GI signal can in principle be removed with a particular linear combination of tomographic shear power spectra~\cite{joachimi08}.

The ``modeling'' technique instead specifies a parameterized model of the intrinsic alignments and then marginalizes over the instrinsic alignment parameters when inferring cosmological parameter constraints ({\it i.e.} the systematic error is subtracted and reduced to a statistical error)~\cite{king02,bridle07}.  This has the advantage of not requiring precise galaxy redshift measurements, but has the disadvantage of being potentially sensitive to the choice of intrinsic alignment model and the number of parameters used in the model (with more parameters potentially leading to larger degradations in the inferred cosmological constraints~\cite{bridle07}).  Recent work has also shown that it may be possible to ``self-calibrate'' the IA signal using the cross-correlation between the ellipticity and galaxy density fields in the same survey~\cite{zhang08}.

Ideally a sufficiently good prediction of the IA signal could be made from simulations and subtracted from the data to leave the pure lensing signal.  However current simulations have some way to go before they will make reliable predictions, due to the finite resolution of a large box size, and the difficulty of including baryonic material. And in any case it will be necessary to have predictions for a range of cosmological models and thus an analytic fitting formula would be very helpful.  The linear alignment model is likely to hold on large scales, whereas it takes no account of non-linear growth of structure. An ad-hoc attempt to rectify this was indicated in Ref.~\cite{hirata07}, and used in~Ref.~\cite{bridle07}, in which the linear theory matter power spectrum in the linear alignment model was replaced by the non-linear matter power spectrum. We refer to this as the nonlinear alignment model (NLA) here.  In this paper we construct an improved model for the IA power spectra based on the halo model.

The halo model of galaxy clustering~\cite{scherrerb91,scoccimarroshj01} has proved a surprisingly successful predictor of galaxy clustering statistics~\citep[see][for a review]{coorays02}.  In this picture the universe consists of dark matter halos that are clustered according to linear theory (in the case of two-point function predictions). Each halo has a mass drawn from a mass function, and a density profile. These are usually taken from average properties of n-body simulations. Galaxy positions are drawn from the resulting dark matter distribution. The two-point correlation function of galaxy positions then constitutes a ``two-halo'' term, arising from the correlations between the positions of two different halos, and a ``one-halo'' term arising from pairs of galaxies that reside in the same halo.

This paper is organized as follows.  We describe our model for central and satellite galaxy ellipticities in Section~\ref{sec:model_for_galaxy_ellipticities}.  In Section~\ref{sec:intrinsic_ellipticity_power_spectra} we derive the 3D auto power spectra of the projected intrinsic galaxy ellipticity distribution and the cross-power
spectra with the matter distribution.  We derive the intrinsic alignment contribution to the angular shear power spectra in Section~\ref{sec:cosmological_implications} and show how constraints on $\sigma_8$ could be biased if the intrinisc alignment correlations were modeled incorrectly.  In Section~\ref{sec:conclusions} we draw conclusions about this new satellite contribution to the intrinsic alignment signal and describe some questions for future numerical simulations to address in order to refine our model. We describe the normalization of the 3D ellipticity power spectra accounting for spatial clustering of the galaxies in Appendix~\ref{sec:marked_power_spectra}
and give details on our analytic model calculation in Appendix~\ref{sec:computation_of_ellipticity_power_spectra}.

Unless otherwise stated, we assume a fiducial cosmological model with $\sigma_8=0.8$, $\Omega_m=0.3$, $\Omega_k=0$, $n_s=1$, and $H_0=71.9$ km s$^{-1}$ Mpc$^{-1}$.

\section{Model for galaxy ellipticities} 
\label{sec:model_for_galaxy_ellipticities}
We are concerned with modeling the correlations in the intrinsic projected orientations of galaxies that could mimic those induced by gravitational lensing.  There are three effects to consider when constructing such a model: the 3-D shapes of galaxies, projection effects, and the relation of the projected ellipticity to the shear defined in gravitational lensing studies.

Elliptical galaxies can generally be described by 3-D ellipsoids, which are further found to be mostly prolate~\cite{allgood06}. Assuming the two minor axes of the ellipsoid have equal lengths on average, we can then approximate elliptical galaxies as sticks corresponding to the length and orientation of the major axis.  In this paper we assume that all halo satellite galaxies can be described in this way.  If the stick makes an angle $\theta$ with respect to the line-of-sight we relate the observable projected intrinsic
ellipticity  of the galaxy to the length of the stick $\ebar$ as,
\begin{equation}
|\intshear| \equiv \frac{a-b}{a+b}
= \ebar \sin\theta.
\end{equation}
where $a$ and $b$ are the major and minor axes of the 2-D projected ellipticity of the galaxy.  We follow the usual convention and define the complex projected ellipticity
\begin{equation}
\intshear(\rv) = |\intshear (r,\theta)|
\, e^{2i\phi} \equiv \eplus(\rv) + i\ecross(\rv).
\end{equation}

\begin{figure}
\centerline{
\includegraphics[width=8cm]{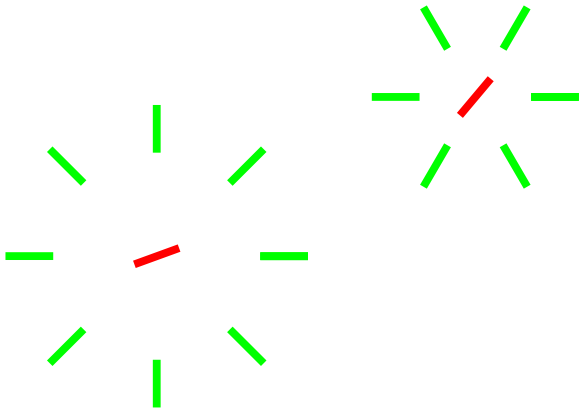}
}
\caption{\label{fg:cartoonmulthalos}Cartoon depiction of the galaxy alignments within and between halos.  The intrinsic alignment correlations within a single halo come from the radial alignment of the satellites (green lines).  The correlations between separate halos are dominated by the correlation of the central galaxies in each halo, depicted as red lines.}
\end{figure}

Figure~\ref{fg:cartoonmulthalos} is a schematic picture of our stick model showing how we populate spherical dark matter halos with central and satellite galaxies.  We describe the separate intrinsic alignment models for centrals and satellites in the next two sub-sections.

\subsection{Satellite galaxies} 
\label{sub:satellite_galaxies}
We build a model for intrinsic alignments of satellite galaxies by populating a spherical dark matter halo with galaxies, with a number density that follow the density profile of the halo.  Our ``basic'' model for satellites assumes that all satellite galaxy ellipticities are aligned with the radius vector the halo, in three dimensions.  This model is inspired by the idea that tidal forces within the halo are responsible for the intrinsic alignments~\cite{pereira08}.

In a flat-sky coordinate system with the origin in the center of the halo, the $z$ axis aligned with the line-of-sight direction, and polar angle $\theta$, the 3-D density-weighted projected ellipticity of galaxies in the halo is (using the same notation as Ref.~\cite{hirata04})
\begin{eqnarray}\label{eq:ellipticitymodel}
\densshear(\rv,m,c) &=&
\intshear(\rv,m,c) \,N_g \, u(\rv|m,c)\\
&=& \ebar(r,m,c)
\, e^{2i\phi}\, \sin\theta\,
 N_g \, u(\rv|m,c)
\end{eqnarray}
where
$\ebar(r,m,c)$
is the magnitude of the projected ellipticity at radius $r$ in a halo of mass $m$, and $c$, which denotes the concentration of the NFW halo profile $\rhonfw$~\cite{nfw}.
We follow Ref.~\cite{hirata04} in working with the density-weighted ellipticity.
In general, $\ebar(r,m,c)$ may be an arbitrary (positive-valued) function of the radial position within and mass of the halo, and may vary from halo to halo.
However, in all the numerical calculations we perform below we set
$\ebar(r,m,c)=0.2$, independent of position, mass and concentration.

\begin{figure}
\centerline{
\includegraphics[width=8cm]{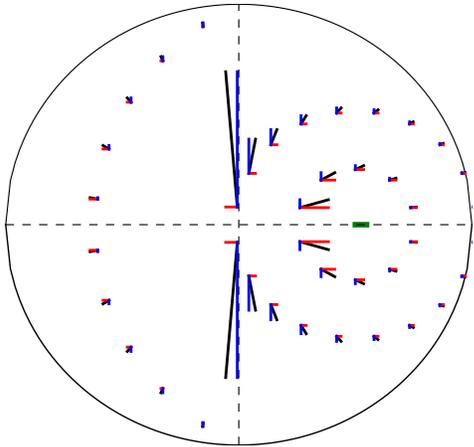}
}
\caption{\label{fg:cartoon}
Depiction of intrinsic alignment correlations of satellite galaxies inside a single halo of mass $10^{14}\,M_{\odot}/h$.  The green line is ellipticity of a typical galaxy that we may be correlating with all the other galaxies in the cluster. The black lines show density weighted ellipticities of other cluster galaxies at 3 fixed radii from the green galaxy.  The red and blue lines are the components of the black lines denoting the degree of correlation with the green galaxy.  Red indicates positive correlation while blue indicates negative correlation.  The total correlation from the 3 circles can be obtained by adding the red and blue lines by eye - which makes it very clear that the smaller circle is strong positive correlation while the middle circle has a weak negative correlation and the largest circle again has a small positive correlation.
}
\end{figure}

\begin{figure}
\centerline{
\includegraphics[width=8cm]{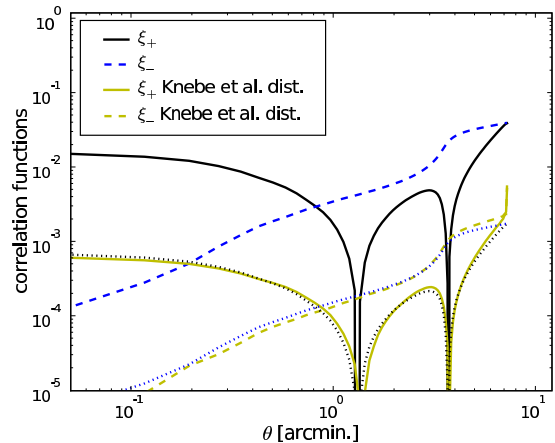}
}
\caption{\label{fg:xialignangledist}Effect of a distribution in radial alignment angles of the satellite galaxy ellipticities on the one-halo correlation function.
The upper lines (black/solid and blue/dashed)  are the two ellipticity correlation functions for a single halo of mass $M=10^{14}\, h^{-1}M_{\odot}$ and concentration $=5$ at $z=0.3$ with perfect radial alignments of the satellites.
(Absolute values are shown.)  In yellow (gray) are the same correlation functions when the radial alignments angles of the satellites  are perturbed by the distribution from Ref.~\cite{knebe08} (also given in Eqns.~\ref{eq:Knebedist} and \ref{eq:etadist}).  The dotted lines show the black and blue upper lines rescaled by the multiplicative factor $\ebareff^2=0.21^2$.  The dotted lines almost perfectly overlay the correlation functions with the distribution in alignment angles.}
\end{figure}

An illustration of the ellipticity correlations induced by this radial stick model inside a single halo is shown in Fig.~\ref{fg:cartoon}. An example galaxy is shown in green in the middle right of the figure, and the ellipticities of three concentric circles of galaxies are shown around this using black lines. Note that these lines vary in length due to the density weighting and there are no galaxies belonging to this halo outside the virial radius, denoted by the black circle. Components of the black lines parallel and perpendicular to the green line are shown in red and blue respectively.  The cartoon shows that this radial alignment model gives a characteristic shape to the one-halo correlation function (as usual $\xi_{\pm}(\theta) = \xi_{++}(\theta) \pm \xi_{\times \times}(\theta)$ where $\xi_{+,\times}(\theta) = \langle \langle \ebar_{+,\times}(\thetav') \ebar_{+,\times}(\thetav'+\thetav) \rangle_{\thetav'} \rangle_{\text{directions of }\thetav}$, where $\ebar_{+}$ and $\ebar_{\times}$ are the components of $\ebar_1$ and $\ebar_2$ in the coordinate system aligned with $\thetav$
-- see also Eqn.~(\ref{eq:normxi}) 
).
The red lines clearly dominate for the central circle, causing a positive correlation. The next circle is dominated by perpendicular components shown in blue, producing a net negative correlation. Finally the largest circle containing any galaxies within the virial radius will produce a positive correlation. See the black line in Fig.~(\ref{fg:xialignangledist}) for an example correlation function (averaged over all positions of the green line in Fig.~\ref{fg:cartoon} as usual).
We computed this correlation function both using the spherical harmonic approximations used throughout this paper and detailed in the Appendices, and also using a Monte Carlo simulation of galaxies placed into an NFW halo. We find excellent agreement.  The one-halo satellite-satellite term in the left panel of Fig.~\ref{fig:Pkcomponents} shows the E-mode power spectrum constructed from the Fourier transforms of $\xi_{\pm}$ in Fig.~\ref{fg:xialignangledist} (according to, e.g., Eqn.~(\ref{eq:normxi})).

\subsection{Central galaxies} 
\label{sub:central_galaxies}
We use the linear alignment model~\citep{catelan01,heavens00,hirata04} to specify the intrinsic ellipticities of central galaxies in our halo model.  Central galaxies are defined such that each dark matter halo has one central galaxy that resides exactly at the center of the spherical halo mass distribution.  We assume that central galaxies have the same ellipticity orientation as their parent halo and that halo ellipticities are determined by the large-scale density perturbations so that \cite[eq.~13 in Ref.~][]{hirata04},
\begin{equation}
\intshear = -\frac{\cla}{4\pi G}
\left(\nabla_{x}^2-\nabla_{y}^2,2\nabla_x\nabla_y\right)\,S[\Psi_P],
\end{equation}
where $S[\Psi_P]$ is the primordial potential with small-scale perturbations smoothed out and $\cla$ is a normalization constant. Unless otherwise stated we take $\cla = 5\times 10^{14}~\left(h^2 M_{\odot}\,\text{Mpc}^3\right)^{-1}$.
This is motivated by comparison with Fig.~2 in Ref.~\cite{hirata04} who normalise to SuperCOSMOS~\cite{heymans04}, and that fits well with the SDSS L4 points of~Ref.~\cite{mandelbaum06} as illustrated in Ref.~\cite{bridle07}.

We can recover the linear alignment model in our framework by setting the Fourier transform of $\intshear(\rv)$ to,
\begin{equation}\label{eq:LA}
\intshear(\kv)=\frac{\cla\,\bar{\rho}}{\bar{D}}
\sin^{2}\theta_k\,\left(\cos(2\phi_k),\sin(2\phi_k)\right)\delta_{\text{lin}}(\kv),
\end{equation}
where $\phi_k$ is the azimuthal angle of
the Fourier wavenumber
$\kv$ about the line-of-sight,
$\theta_k$ is the polar angle of $\kv$ with respect to the line of sight, $\delta_{\text{lin}}$ is the linear theory density perturbation and the remaining definitions follow Ref.~\cite{hirata04}.



\section{Intrinsic ellipticity power spectra} 
\label{sec:intrinsic_ellipticity_power_spectra}
We construct a continuous intrinsic ellipticity field by following the usual halo model assumption~(e.g. Ref.~\cite{cooray02}) in which we sum the density-weighted ellipticity distribution for a single halo over halo masses $m_i$ and positions $\rv_i$,
\begin{align}\label{eq:densshearHM}
\densshear(\rv) &= \frac{1}{\ngbar}\sum_{i}
\intshear(\rv-\rv_i,m_i)\,N_{g,i}\,
u(\rv-\rv_i|m_i)
\notag\\
&= \sum_i \int dm \int d^{3}r'
\delta\left(m-m_i\right)\delta^{(3)}\left(\rv'-\rv_i\right)\,
\notag\\
&\qquad\times\frac{N_{g,i}}{\ngbar}
\intshear(\rv-\rv',m) u\left(\rv-\rv'\right|m),
\end{align}
where $u(\rv|m)\equiv\rhonfw(\rv,m)/m$, $N_{g,i}$ is the number of galaxies in the $i$th halo, and $\ngbar$ is the mean number of galaxies per unit volume (at a given redshift).
We will assume a deterministic relation between halo mass $m$ and concentration $c$ hereafter with $c(M,z)=\left(9/(1+z)\right) \left(M/M_{*}(z)\right)^{-0.13}$~\cite{cooray02}.

We then compute the 3-D ellipticity power spectra by first Fourier transforming each component of Eqn.~(\ref{eq:densshearHM}),
\begin{equation}
\densshear_{j}
(\kv,m) \equiv
\int d^3\rv\,
\densshear_{j}
(\rv,m)\, e^{i\kv\cdot\rv}
\end{equation}
where $j=1,2$ denotes the shear component. We perform this Fourier transform for our model using a multipole expansion of the plane waves as shown in Appendix~\ref{sub:satellite_ellipticity_density_run}.

\begin{figure*}
\includegraphics[width=8cm]{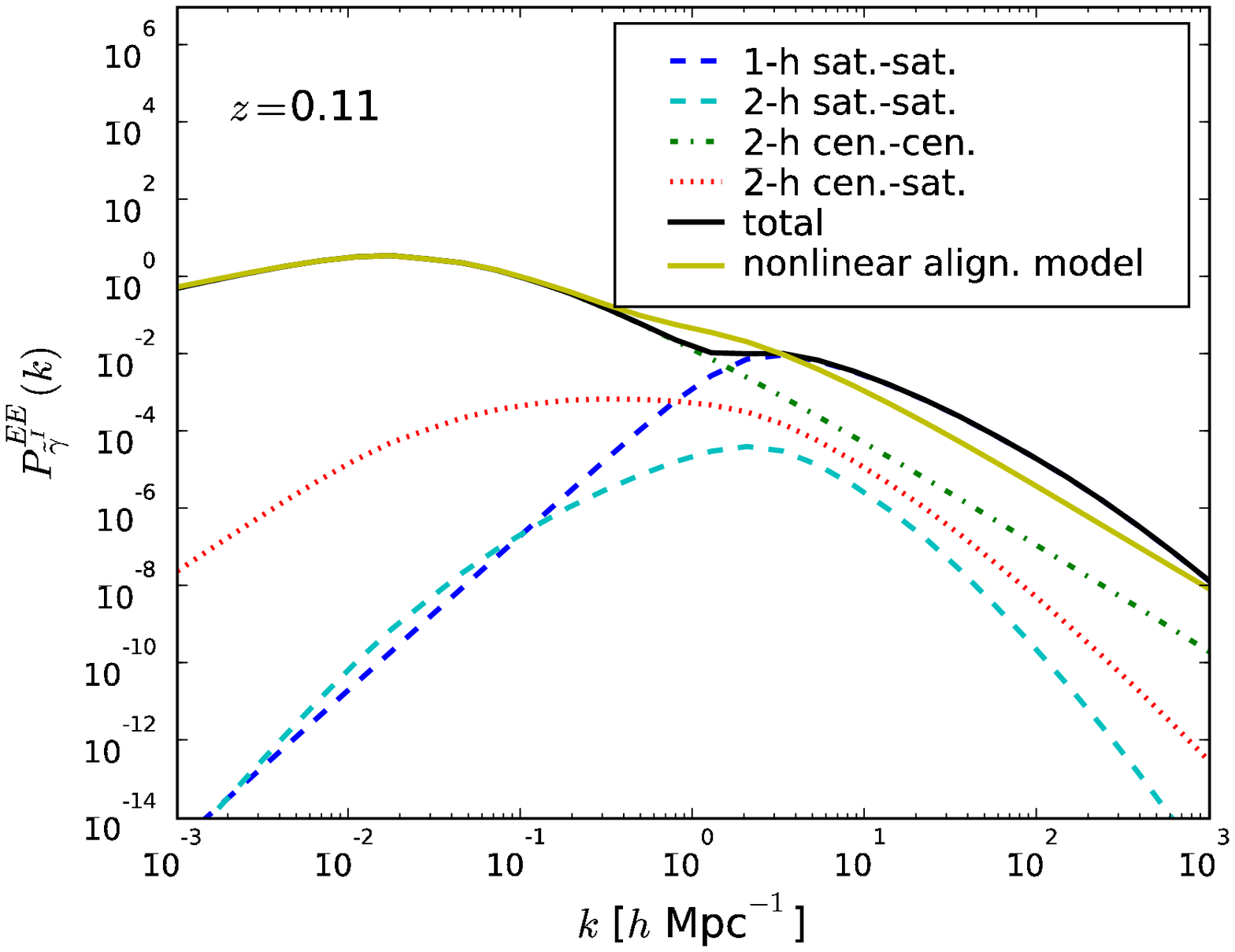}
  \label{fig:IIEE}
\includegraphics[width=8cm]{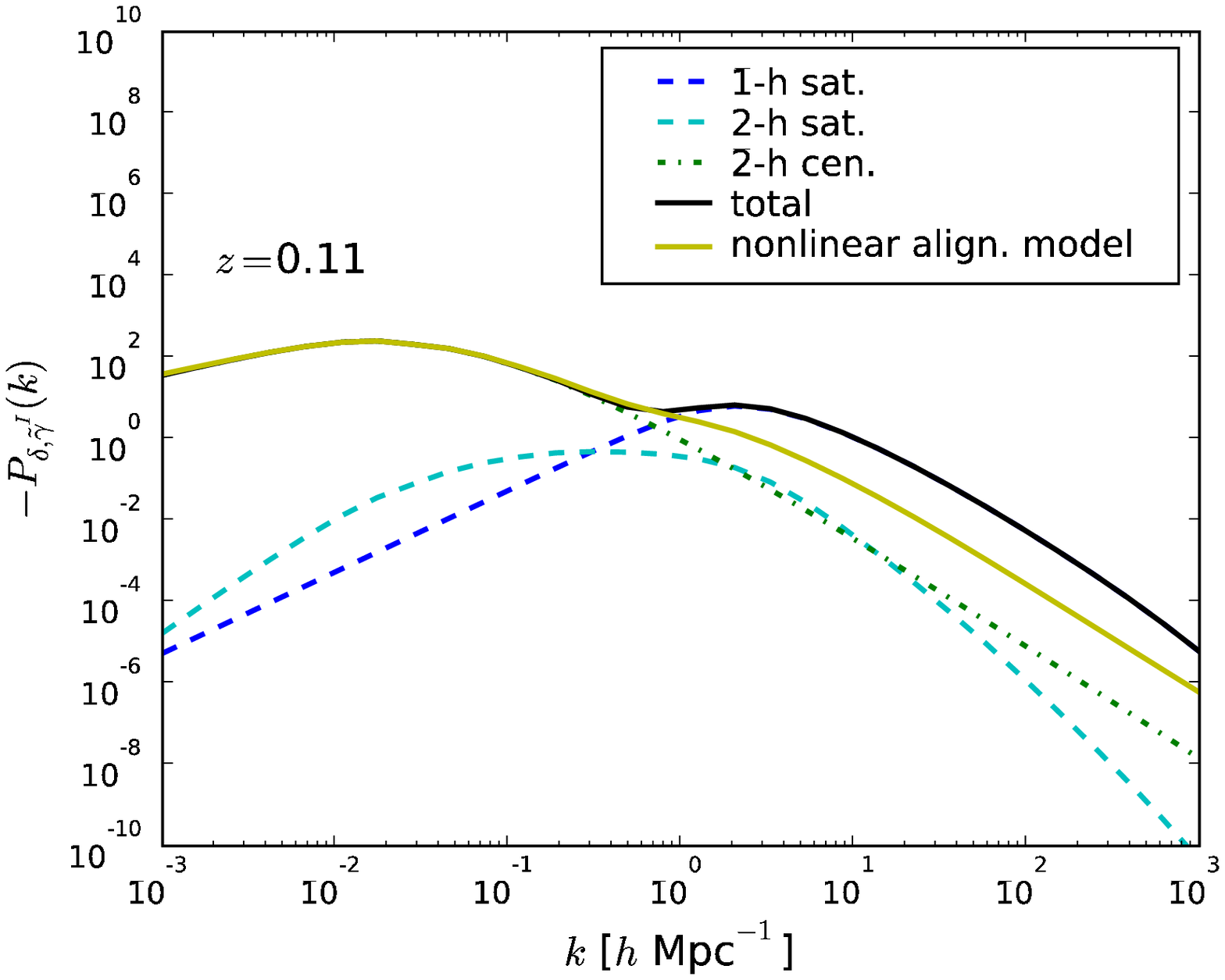}
  \label{fig:GIEE}
\caption{\label{fig:Pkcomponents}All nonzero contributions to the II and GI E-mode power spectra in the spherical, radially aligned satellite, halo model.
For comparison we plot the ``nonlinear alignment model'' in which the non-linear matter power spectrum is used within the linear alignment model.}
\end{figure*}

The E and B Fourier modes are defined as
\begin{align}\label{eq:EBmodes}
\densshear_{E}(\kv) &= \cos(2\phi_k)
\densshear_{1}(\kv)
+\sin(2\phi_k)
\densshear_{2} (\kv)
\notag\\
\densshear_{B}(\kv) &= \sin(2\phi_k)
\densshear_{1} (\kv)
-\cos(2\phi_k)
\densshear_{2}
(\kv).
\end{align}
We then define the 3-D ellipticity power spectra,
\begin{align}\label{eq:EBspectraDef}
\left< \tilde{\gamma}^{I*}_{E} (\kv)\densshear_{E}(\kv')\right> &=
(2\pi)^3\delta^{(3)}_{D}\left(\kv-\kv'\right) P^{EE}_{\densshear}(\kv)
\notag\\
\left< \tilde{\gamma}^{I*}_{B} (\kv)\densshear_{B}(\kv')\right> &=
(2\pi)^3\delta^{(3)}_{D}\left(\kv-\kv'\right) P^{BB}_{\densshear}(\kv)
\notag\\
\left< \delta^{*}(\kv)\densshear_{E}(\kv')\right> &=
(2\pi)^3\delta^{(3)}_{D}\left(\kv-\kv'\right) P_{\delta,\densshear}(\kv).
\end{align}
When considered as contributions to the cosmic shear signal, $P^{XX}_{\densshear}$ gives the II contribution from the intrinsic-ellipticity correlation of two galaxies  in the lens plane, while $P_{\delta,\densshear}$ gives the GI contribution from the correlation of a galaxy in the lens plane with its surrounding dark matter distribution (which is the lens).

First we consider the case where all the satellite galaxies are pointing directly towards the center of the halo. Then we consider the case where there is some randomness in this direction. We then present fitting formulae for the above power spectra.

\subsection{Radially aligned satellites}
\label{sec:radially_aligned_satellites}

Two types of term appear when taking products of the form in Eqn.~(\ref{eq:EBspectraDef}): both galaxies are in the same dark matter halo or they are in different halos.  These are commonly referred to as the ``one-halo'' and ``two-halo'' contributions to the power spectrum.  In addition, because we have separate models for central and satellite galaxies in halo (and because the probabilities for centrals and satellites to occupy a halo of mass $m$ may be different) we get an additional separation of terms into correlations of satellites with satellites, centrals with centrals, and satellites with centrals.

By integrating over the joint probablitity distribution for 2 halos of masses $m_1$ and $m_2$ to be at positions $\rv_1$ and $\rv_2$ with $N_{g,1}$ and $N_{g,2}$ galaxies in each halo we arrive at the satellite-satellite 3-D II power spectra,
\begin{align}\label{eq:3dspectrum}
P^{EE,1h}_{\densshear,ss} &= \int dm\, n(m)\,
\frac{\left<N_g^s(N_g^s-1)|m\right>}{\ngbar^2}\,
\notag\\
&\times\ebar^2(m)\,
\left| w(k,\theta_k|m)\right|^2
\notag\\
P_{\densshear,ss}^{EE,2h}(k) &= \int dm_1\, n(m_1)\,
\frac{\left<N_g^s|m_1\right>}{\ngbar}\,\ebar(m_1)\,
\left| w(k,\theta_k|m_1) \right|
\notag\\
&\times
\int dm_2\, n(m_2)\,
\frac{\left<N_g^s|m_2\right>}{\ngbar}\,\ebar(m_2)\,
\left| w(k,\theta_k|m_2) \right|
\notag\\
&\times P_{hh}(k|m_1,m_2),
\end{align}
where $P_{hh}$ is the halo-halo power spectrum, $w(k|m)\equiv \densshear(\kv,m) / \ebar(m)$ (see Eqn.~(\ref{eq:densRunDefinition})), $\left<N_g^s|m\right>$ and $\left<N_g^s(N_g^s-1)|m\right>$ are the first and second moments of the galaxy number distribution within a halo of mass $m$, and the $s$ and $c$ subscripts denote ``satellite'' and ``central'' galaxies, respectively.

The two-halo central-central power spectrum is (Eqn.~(16) in~\cite{hirata04}),
\begin{align}
P_{\densshear,cc}^{EE,2h}\left(k\right) &=
\cla^2\, P_{\text{lin}}(k) +
\notag\\
&\cla^2\, b_{g,c}^2\int \,\frac{d^3k_1}{(2\pi)^3} \left[f_{E}(\kv_1)+
f_{E}(\kv_2)\right]\,
\notag\\
&\times f_{E}(\kv_1)\,
P_{\text{lin}}(k_1)\,P_{\text{lin}}(k_2),
\end{align}
where $\kv_2\equiv \kv-\kv_1$, $C\equiv \cla\bar{\rho}/\bar{D}$, $b_{g,c}$ is the linear galaxy bias for the central galaxy population, and, following Ref.~\cite{hirata04}, $f_{E}(\kv)\equiv \left(k_x^2-k_y^2\right)/k^2$.
The two-halo central-satellite term in the power spectrum is,
\begin{align}
P_{\densshear,cs}^{EE,2h}(k) &= \frac{\cla \bar{\rho}}{\bar{D}}\,
P_{\text{lin}}(k)\,
\notag\\
&\times\int dm\, n(m) \frac{\left<N_g^s|m\right>}{\ngbar} b_h(m)
\notag\\
&\times
\ebar(m)\, \left|w(k,\theta_k|m)\right|,
\end{align}
where $b_h$ is the halo bias.
In principle there could be a one-halo central-satellite correlation, but this is zero when using the linear alignment model for the central galaxies and assuming that the satellite galaxy ellipticities are uncorrelated with the central galaxy ellipticity.

Because we have normalized by the comoving density of galaxies $\ngbar(z)$, these power spectra are only very weakly dependent on the model redshift distribution.  We have verified that the weight in the II integrand, $m\,n(m,z)\,\left<N_g|m\right>/\ngbar(z)$ has an r.m.s. error of less than 10\% between redshift distributions with median $z\sim0.6$ and $\sim1.7$.
Therefore we consider only a single redshift distribution for the remainder of this paper (which is the model from Ref.~\cite{song04} with limiting magnitude in {\it R} of 26.).
Changing the relative contribution of spiral and elliptical morphologies to the satellite population, and assuming that spirals have zero intrinsic alignments, does affect the power spectrum by diluting the IA signal.  Using ad-hoc models for the halo mass-dependent ratio of spiral and elliptical satellites we find the dominate effect is to simply rescale the amplitude of the power spectra. 

We show in Appendix~\ref{sub:satellite_ellipticity_density_run} that $\densshear_{B,s}=0$; implying $P^{BB,1h}_{\densshear,ss} = P^{BB,2h}_{\densshear,ss} = P^{BB,2h}_{\densshear,cs}=0$.  This follows directly from the simple, and separable, $\phi$-dependence in $\intshear(\rv)$.  We can understand this more directly by noting that all the satellite galaxy ellipticities are perpendicular to the boundary of the halo (assumed to be truncated at the virial radius).  From, e.g., Ref.~\cite{bunn03} we know that this boundary condition is incompatible with a nonzero $B$-mode\footnote{Although, this condition does not distinguish ``pure'' E modes from ambiguous modes as defined in Ref.~\cite{bunn03}.  But, this distinction is not important for this simple model.}.  The linear alignment model B-mode, $P^{BB,2h}_{\densshear,cc}$ was derived in Ref.~\cite{hirata04}, but is much smaller than the E-mode and we do not consider it further.

Some of the two-halo terms we have to consider for an ensemble of halos are demonstrated in cartoon form in Fig.~\ref{fg:cartoonmulthalos}.
The dominant contribution to the two-halo term is the correlation of the central galaxies.  This makes sense intuitively if we think of averaging the satellite ellipticity correlations by eye in Fig.~\ref{fg:cartoonmulthalos}.

The nonzero GI power spectra are
\begin{align}
P_{\delta,\densshear,s}^{1h} &= \int dm\, n(m)\, \frac{m}{\rhobar}
\frac{\left<N_g|m\right>}{\ngbar}\,
\notag\\
&\times\ebar(m)\,
\left| w(k,\theta_k|m)\right|\,u(k|m)
\notag\\
P_{\delta,\densshear,s}^{2h} &= \int dm_1\, n(m_1)\,
\frac{\left<N_g^s|m_1\right>}{\ngbar}\,\ebar(m_1)\,
\left| w(k,\theta_k|m_1) \right|
\notag\\
&\times
\int dm_2\, n(m_2)\,
\frac{m}{\rhobar}\,u(k|m_2)
\notag\\
&\times P_{hh}(k|m_1,m_2)
\notag\\
P_{\delta,\densshear,c}^{2h} &= -\frac{\cla \bar{\rho}}{\bar{D}}\,
P_{\text{lin}}(k),
\end{align}
where $P_{\delta,\densshear,c}^{2h}$ is the same as Eqn.~(18) in Ref.~\cite{hirata04}.
We assume that the halo-halo power spectrum $P_{hh}$ is given by the linear theory power spectrum $P_{\text{lin}}$ multiplied by the halo bias for each mass. 
Again, there could be a one-halo central galaxy term, but it is zero for the assumptions we have made in this halo model.

We show the various one-halo and two-halo components of the E-mode II and GI power spectra in Fig.~\ref{fig:Pkcomponents}.  The total power spectra are well-described by only considering the satellite contribution to the one-halo term and the central contribution to the two-halo term.  We will neglect the other cross-terms shown in Fig.~\ref{fig:Pkcomponents} in our remaining analysis.  We see that the one-halo contributions to both the II power spectrum becomes important around $k=1 h$ Mpc$^{-1}$. This fits quite well with the increase in power in the non-linear matter power spectrum.
The amplitude on small scales is similar to the nonlinear alignment model for the II power spectrum, and significantly larger for the GI power spectrum.

\subsection{Distribution in radial alignment angles} 
\label{sub:distribution_in_radial_alignment_angles}

In this section we consider the effect of a distribution in the angle between the galaxy major axis and the radial vector for satellites in a halo.  If $\hat{r}$ is a unit radial vector in the halo and $\hat{e}$ is a unit vector denoting the orientation of the galaxy major axis, then we consider a distribution in the angle $\beta$ where,
\begin{equation}
\cos\beta \equiv \hat{r}\cdot\hat{e}.
\end{equation}

By default, we adopt the distribution found in Ref.~\cite{knebe08} (their Eqn.~(5)) from N-body simulations,
\begin{equation}\label{eq:Knebedist}
P(\beta) = \sin(\beta)\left(\frac{A}{B+A/5}\cos^4\beta+B\right)
\end{equation}
with $A=2.6$ and $B=0.6$.
Once we allow a misalignment between the galaxy major axis and the halo radial vector then there is a second angle to consider that denotes rotations about the axis defined by $\hat{r}$, which we label $\eta$.  We assume a uniform distribution,
\begin{equation}\label{eq:etadist}
P(\eta) = \frac{1}{2\pi}.
\end{equation}
We derive the satellite ellipticity as a function of $\beta$ and $\eta$ in Section~\ref{sub:galaxy_major_axis_as_a_function_of_radial_alignment_angles}.
We have also performed a Monte Carlo simulation of a single halo
to more robustly study the effect of this distribution in alignment angles on the one-halo correlation functions.

\begin{figure}
\centerline{
\includegraphics[width=8cm]{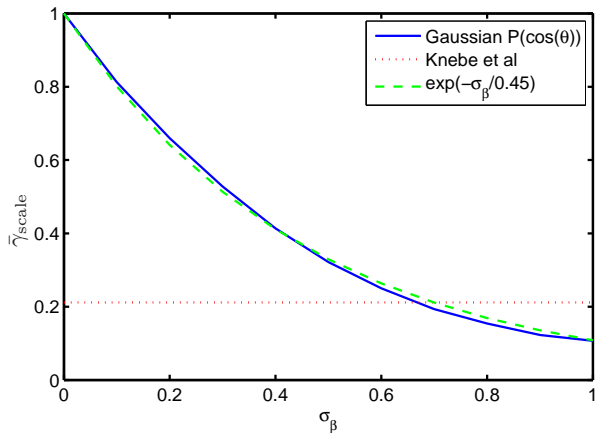}
}
\caption{\label{fg:alignangledist}
Ellipticity suppression factor $\ebareff$ for different levels of random perturbations $\beta$ in the radial satellite alignments.
$\ebareff$ is shown as a function of the Gaussian width of $P(\cos\beta)$ with the result from Ref.~\citep{knebe08} overlaid.}
\end{figure}

The resulting correlation function is shown by the lower pair of lines in Fig.~(\ref{fg:xialignangledist}).  We find the main effect of adding a distribution in radial alignments is simply to reduce the amplitude of the correlation functions with respect to the case with perfect radial alignemnts.  For the distribution in Eqn.~(\ref{eq:Knebedist}), we get a multiplicative amplitude reduction of
\begin{equation}\label{eq:ebardef}
\ebareff^2\equiv 0.21^2
\end{equation}
independent of halo mass.  The dot-dashed lines in Fig.~(\ref{fg:xialignangledist}) show the correlation function without a distribution in alignment angles, scaled down by this factor. It is in good agreement with the full calculation.  Unsurprisingly this suppression factor can also be obtained more quickly by integrating over $\beta$ and $\nu$ for a satellite at a given polar angle $\theta$ in the halo.

We have also performed simulations using a Gaussian distribution in the radial alignment angle $\beta$ with zero mean and a varying width $\sigma_{\beta}$.
The amplitude reduction as a function of $\sigma_{\beta}$ is shown in Fig.~(\ref{fg:alignangledist}) where we find that $\sigma_{\beta}\sim 0.7$ gives similar correlation functions to those obtained using the distribution in Eqn.~(\ref{eq:Knebedist}).

When $P(\beta)$ has finite width, a nonzero B-mode can be generated for $\densshear_{s}$.  We have computed this amplitude via numerical integration of $f_{\ell}$ in Eqn.~(\ref{eq:multipoleexpansion}) (and using Eqn.~\ref{eq:thetae_phie}) with a Gaussian distribution for $P(\beta)$ and find the B-mode amplitude is several orders of magnitude smaller than the E-mode amplitude independent of the width of the $P(\beta)$ distribution.  We therefore ignore the B-mode power spectra in the following results.

\subsection{Fitting functions} 
\label{sub:fitting_functions}

\begin{figure*}
\centering
\includegraphics[width=8.5cm]{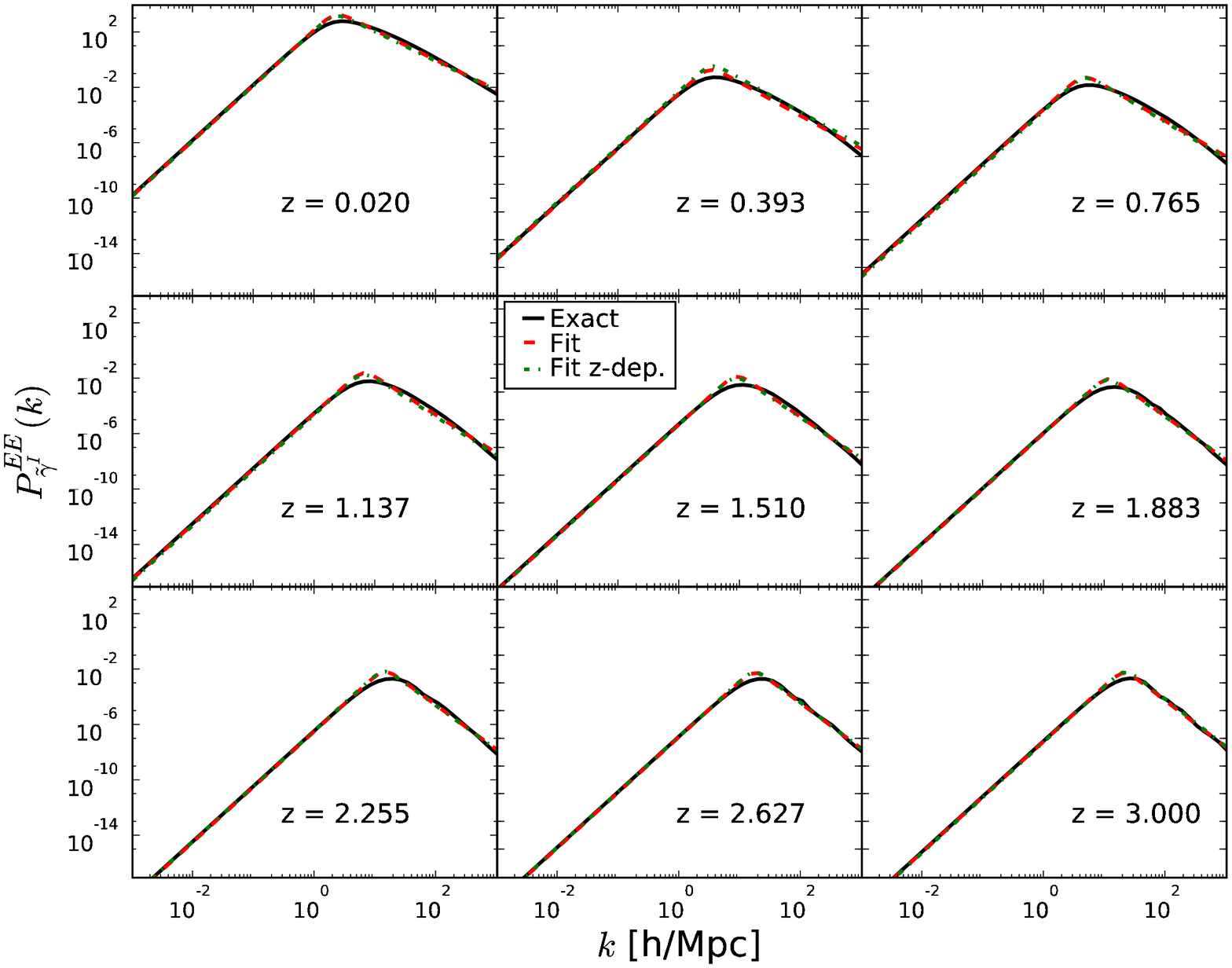}
\includegraphics[width=8.5cm]{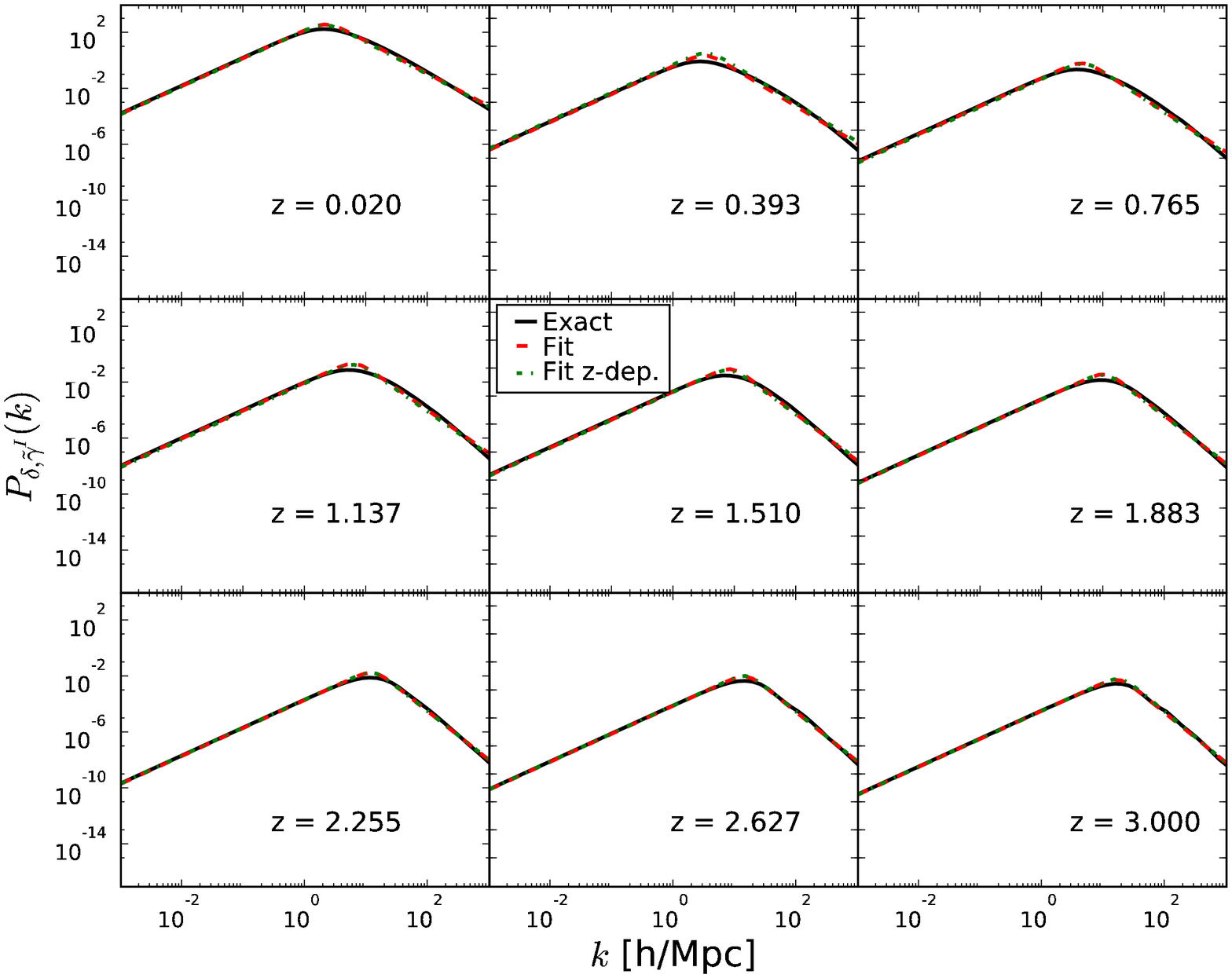}
\caption{\label{fg:fitPk}Comparison of the II and GI one-halo power spectra and their fitting functions at 9 different redshifts.  The dashed (red) lines show the best fit of Eqn.~(\ref{eq:Pkfitfcns}), while the dot-dashed (green) lines show the best fits of the z-dependent function in Eqn.~(\ref{eq:Pkzdepfit}).}
\end{figure*}

\begin{figure*}
\centering
\includegraphics[width=8.5cm]{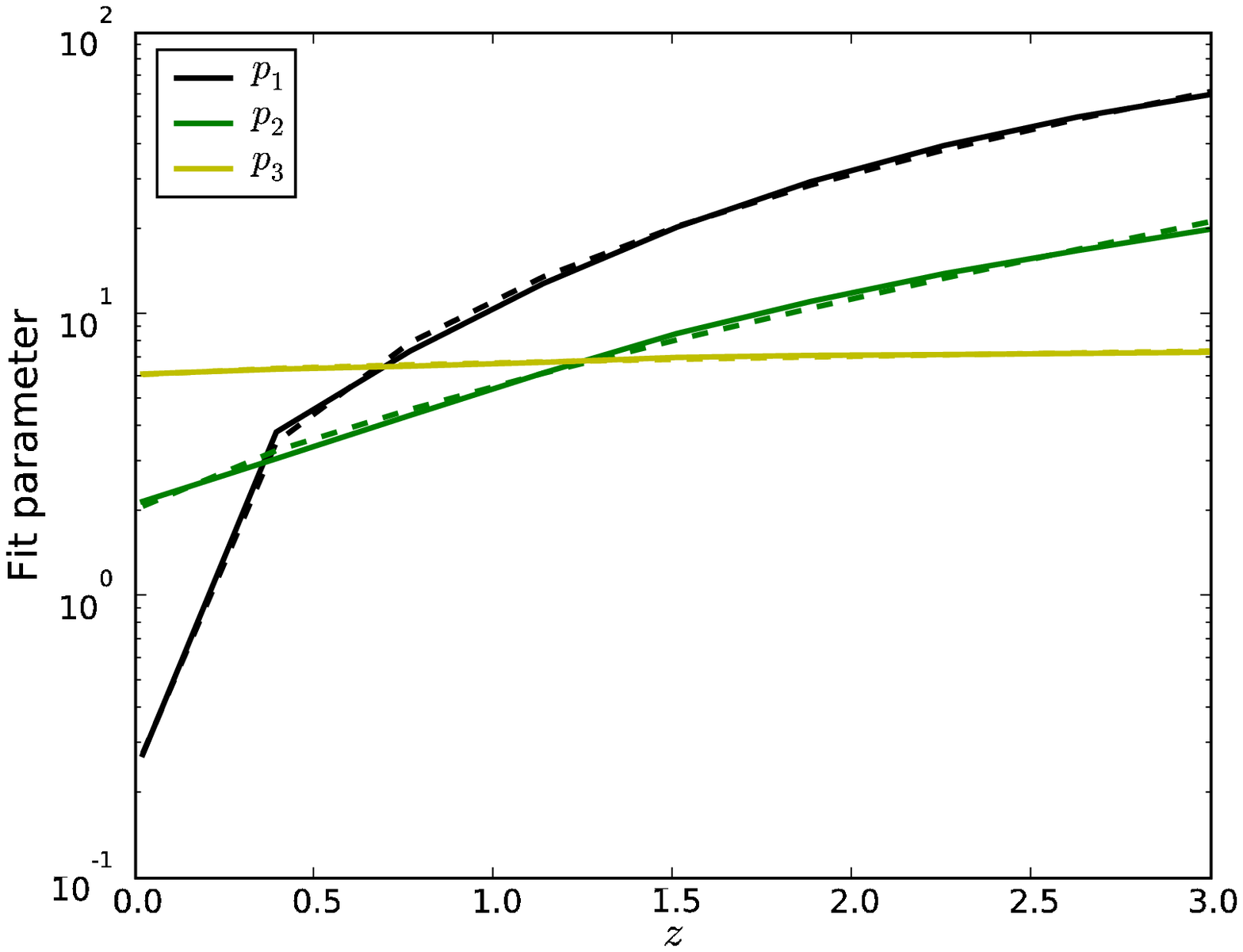}
\includegraphics[width=8.5cm]{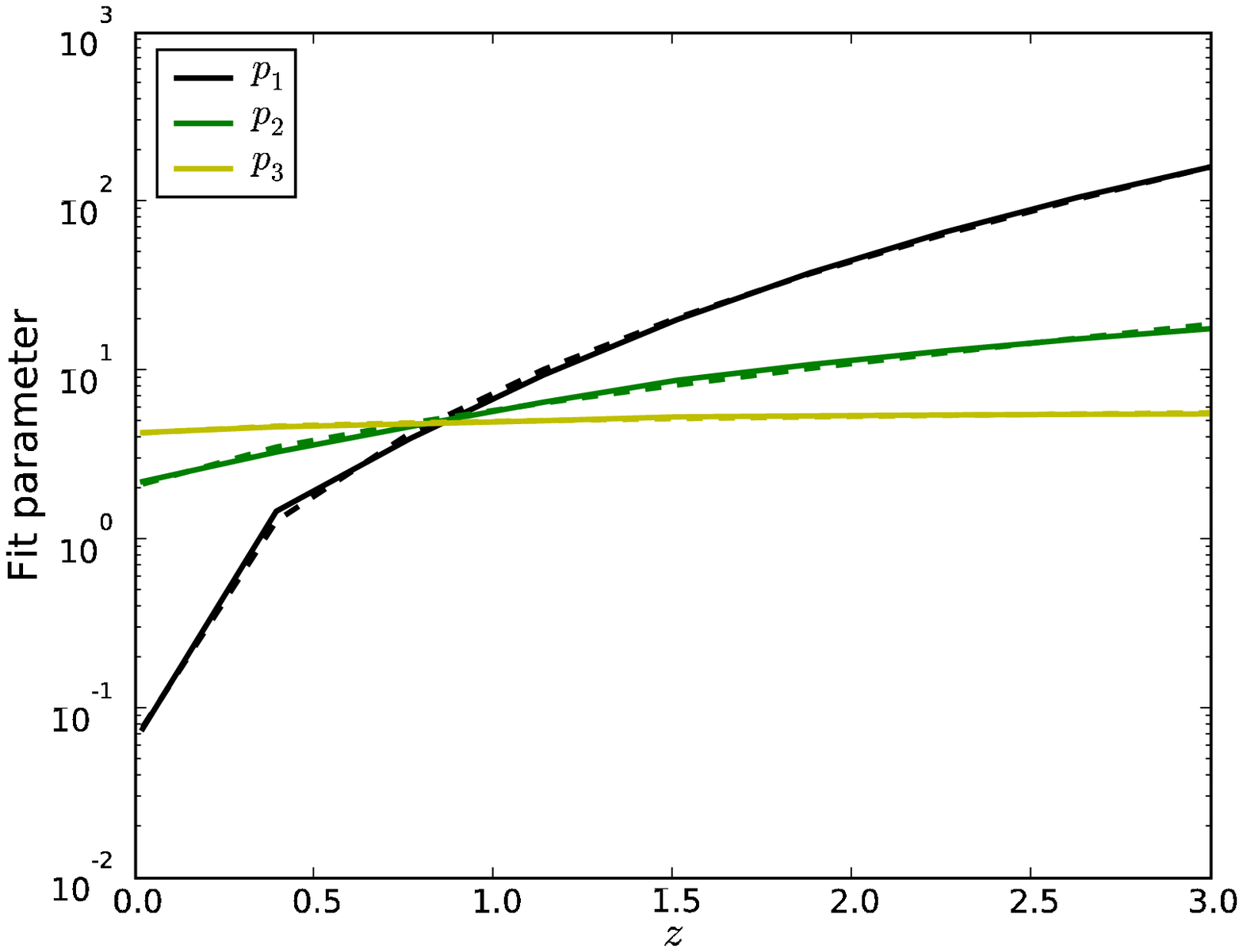}
\caption{\label{fg:fitparams}II and GI power spectra fit parameters as functions of   redshift.  The solid lines are the actual parameters fit at 9 different redshifts and the   dashed lines are 3 parameter fits $q_{ij}$ that we use to model the redshift evolution of the power spectra.}
\end{figure*}

We show the II and GI satellite power spectra at several redshifts in Fig.~\ref{fg:fitPk}
with \eeff set to unity.  The amplitudes of both spectra decrease with increasing redshift, which is largely a consequence of the decreasing fraction of satellite versus central galaxies as the redshift increases (so the one-halo signal is essentially diluted).
The peaks of the spectra also shift to smaller scales with increasing redshift as 
the objects move further away. 

We find the one-halo II and GI power spectra at a given redshift can be reasonably fit with the following 3-parameter functions,
\begin{align}\label{eq:Pkfitfcns}
P^{1h}_{\densshear,ss,\text{fit}}(k) &=
\ebareff^2
\frac{\left(k/p_1\right)^4}
{1+\left(k/p_2\right)^{p_3} }
\notag\\
P^{1h}_{\delta,\densshear,s,\text{fit}}(k) &=
-\ebareff
 \frac{\left(k/p_1\right)^2}
{1 + \left(k/p_2\right)^{p_3} }.
\end{align}

We show the evolution of $p_i$ as a function of redshift in Fig.~\ref{fg:fitparams}.
We fit each $p_i$ (for $i=1,2,3$) for both II and GI fitting functions with the model,
\begin{equation}\label{eq:Pkzdepfit}
p_{i}(z)
= q_{i1}\,\exp\left(q_{i2}\,z^{q_{i3}}\right).
\end{equation}
The best fit values for the $q_{ij}$ are given in Table~\ref{tb:PkIIfitparamsfit} for the II spectra and in table~\ref{tb:PkGIfitparamsfit} for the GI spectra (note that $q_{i1}$ has units of $h$\Mpc$^{-1}$).  The fit functions are compared to the input spectra in fig.~\ref{fg:fitparams}.
\input{Pkfitparamsfit.tex}
\input{PkGIfitparamsfit.tex}

We then have 9 parameters each to describe the redshift dependent II and GI one-halo power spectra, since $q_{11}$ is degenerate with \eeff (as well as the fraction of spiral to elliptical galaxies in a halo as mentioned in Section~\ref{sec:intrinsic_ellipticity_power_spectra}). 
Combined with the free amplitudes for the two-halo terms, we have 20 parameters in total to describe the intrinsic alignment power spectra.



\section{Cosmological implications} 
\label{sec:cosmological_implications}

\begin{figure*}
\centering
\includegraphics[width=16cm]{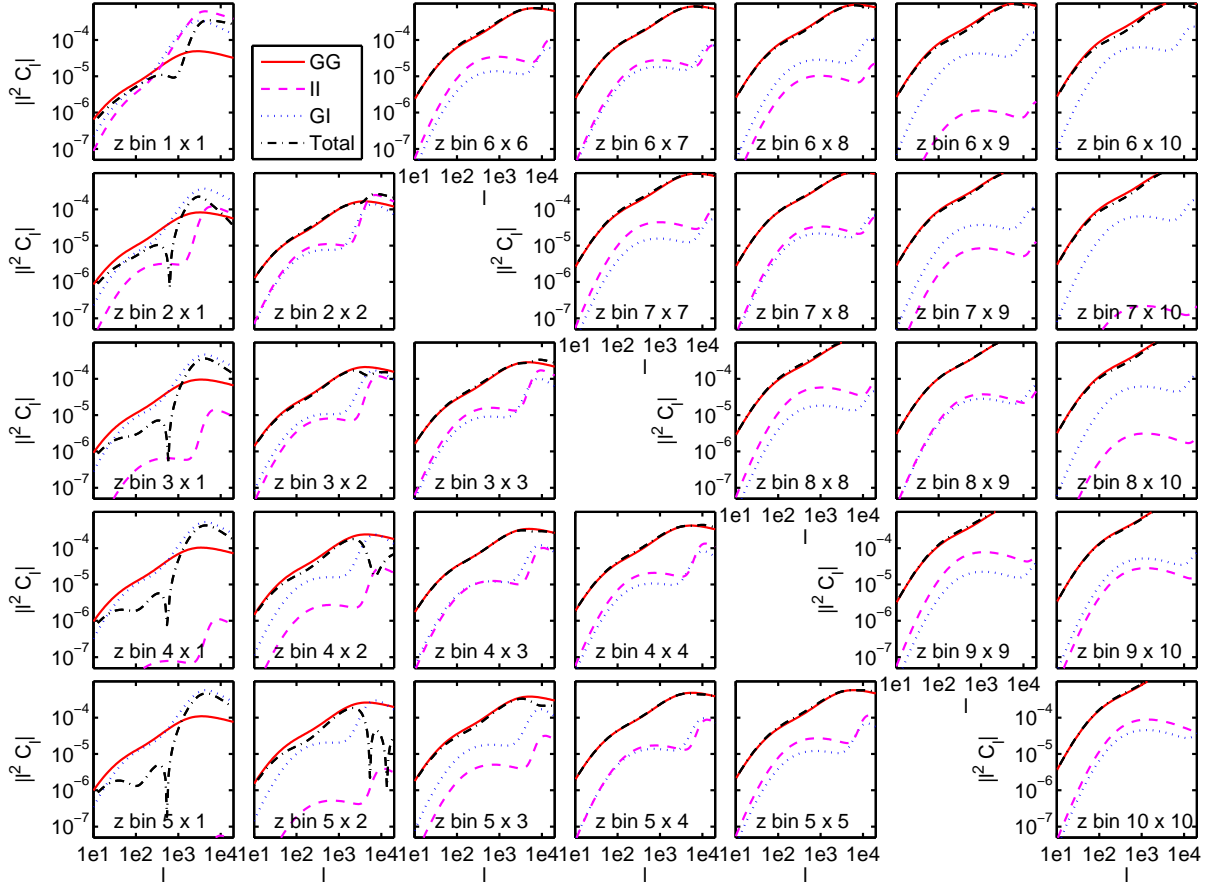}
\caption{\label{fg:cls_tomographic}Angular shear power spectra for a fiducial deep survey with 10 tomographic redshift bins and a Gaussian photometric redshift uncertainty $0.05 (1+z)$.  The solid line shows the usual power spectra from lensing alone (everywhere positive). The dashed lines show the II contribution (everywhere positive). The dotted lines show the GI effect (everywhere negative) and the dot-dashed lines show the total (sum), which can in general oscillate positive and negative depending on the balance of competition between the three effects.
}
\end{figure*}

We use our model to compute all contributions to a set of
angular cosmic shear power spectra
with redshift distribution
\begin{equation}\label{eq:nofz}
\frac{dN}{dzd\Omega}(z) = z^{a}\,\exp\left[\left(\frac{z}{z_0}\right)^{b}\right]
\end{equation}
\citep{smailef94}.

We consider the amplitude of the two-halo term, and the amplitude of the one-halo term as parameterized by the ellipticity suppression factor, \eeff~(Eqn.~(\ref{eq:ebardef})), that we use a proxy for the degree of radial alignment of the satellites in a halo. By default we assume the ``true'' IA signal is given by the halo model with fiducial parameter \eeff $=0.21$.

\subsection{Angular shear power spectra} 
\label{sub:angular_shear_power_spectra}

\begin{figure*}
\centering
\includegraphics[width=15cm]{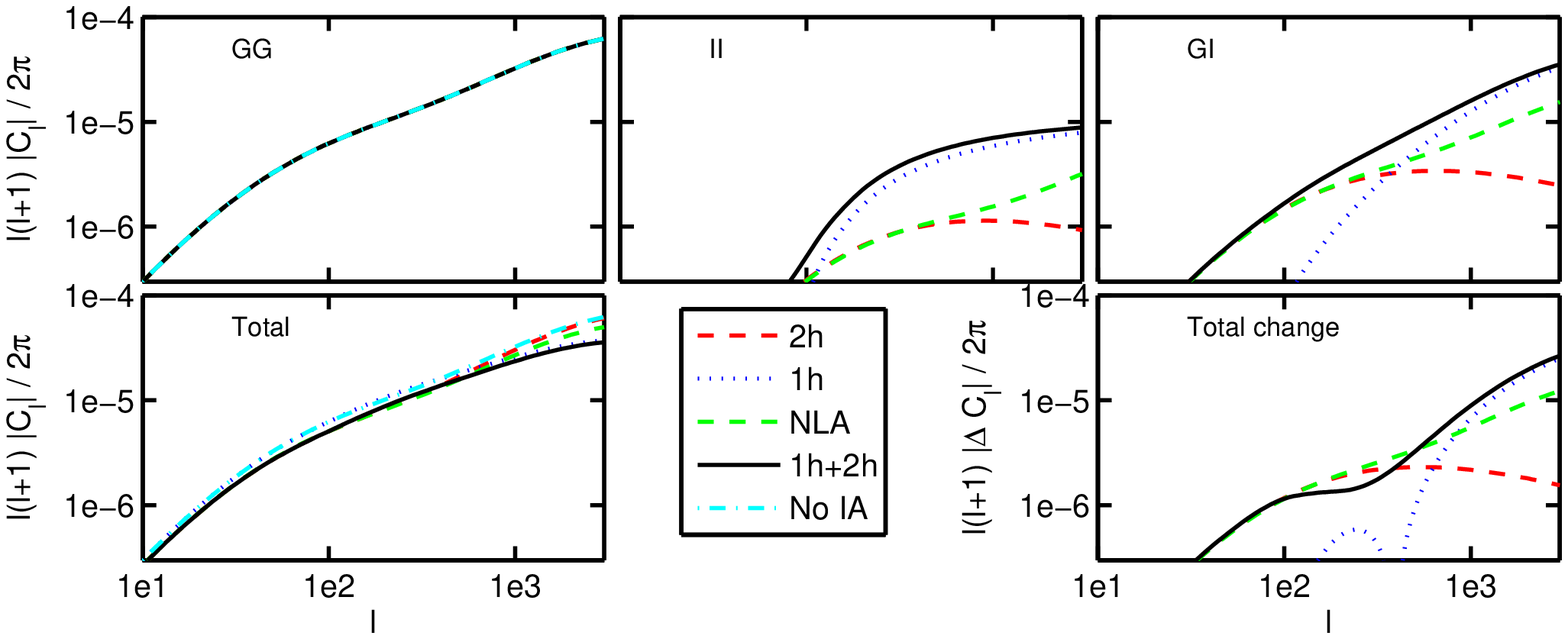}
\caption{\label{fg:clbin_comparison}Angular shear power spectra showing contributions from GG, GI, and II terms for a range of different intrinsic alignment models for a fiducial medium-deep survey.
The dashed line shows the power spectra for the linear alignment model, or equivalently the two-halo term.
The dotted line shows the one-halo term only.
In light/green dashed is the linear alignment model using the non-linear matter power spectrum in place of the linear matter power spectrum. The solid line is the model proposed in this paper, consisting of the one and two halo terms. The power spectra in the absence of intrinsic alignments is shown by the dot-dashed line.
The bottom right panel shows the difference between the total angular power spectra for the various models, as compared to the no intrinsic alignments case.}
\end{figure*}

For comparison with weak lensing power spectra, we need to integrate the
3-D projected ellipticity model over the lensing source distribution
\begin{equation}\label{eq:epsprojection}
\intshear_{\alpha}(\thetav) = \int d\chi\,f_{\alpha}(\chi)
 \densshear(\chi\thetav,\chi),
\end{equation}
where $f_{\alpha}(\chi)$ is the number of sources in sample $\alpha$ per unit comoving distance normalised so $\int f_{\alpha}(\chi) d \chi = 1$.
Therefore, assuming Limber's approximation, the shear angular power spectra can be written
\begin{equation}\label{eq:clsum}
C_{\ell(\alpha \beta)} = C^{GG}_{\ell (\alpha \beta)} +
C^{II}_{\ell(\alpha \beta)} + C^{GI}_{\ell(\alpha \beta)}
\end{equation}
where the terms are sourced by the lensing distortions due the the matter power spectrum and the intrinsic alignment terms calculated in the previous sections
\begin{eqnarray}
\label{eq:cls_GG}
C^{GG}_{\ell(\alpha \beta)}
&=&\!\! \int_0^\infty \!{
q_{\alpha}(\chi) q_{\beta}(\chi)
\over\chi^2}
P_\delta(k;\chi)
d\chi
\nonumber
\\ \label{eq:cls_GI}
C^{GI}_{\ell(\alpha \beta)}
&=&\!\! \int_0^\infty \!{
(q_{\alpha}(\chi) f_{\beta}(\chi) + f_{\alpha}(\chi) q_{\beta}(\chi))
\over\chi^2}
P_{\delta,\tilde\gamma^I}(k;\chi)
d\chi
\nonumber
\\ \label{eq:cls_II}
C^{II}_{\ell(\alpha \beta)}
&=&\!\! \int_0^\infty \!{
f_{\alpha}(\chi) f_{\beta}(\chi)
\over \chi^2}
P_{\tilde\gamma^I}(k;\chi)
d\chi
\end{eqnarray}
where $k={\ell/\chi}$.

We first consider a relatively deep survey with $a=2,\, b=1.5,\, z_0=0.64$ with 10 tomographic redshift bins and a photometric redshift uncertainty of $0.05(1+z)$.
This is more relevant for a surveys such as LSST\footnote{www.lsst.org}, Euclid\footnote{dune-mission.net}
or JDEM\footnote{jdem.gsfc.nasa.gov}. 
A subset of the resulting angular power spectra are shown in Fig.~\ref{fg:cls_tomographic}.
The one-halo contribution is clearly seen as the sharp rise in power in II and GI at small scales.
For this photometric redshift uncertainty the II term is quite comparable or subdominant to the GI term, even in the autocorrelation bins.
The effect of intrinsic alignments is clearly largest at very low redshift, as expected.
There is a competition between the GI, II and GG terms that causes the total power spectra for the lowest bins to oscillate above and below zero (only the absolute magnitudes are shown, but the sign can be identified by recalling that the II and GG power spectra are positive and the GI power spectrum negative).

Fig.~\ref{fg:clbin_comparison} shows the angular power spectra
for a CFTHLS-like survey\footnote{www.cfht.hawaii.edu/Science/CFHLS/}
with $a=0.836,\, b=3.425,\, z_0=1.171$ in Eq.~\ref{eq:nofz}
(taken from Ref.~\cite{benjamin07}).
For comparison with existing results we use a single redshift bin and a photometric redshift uncertainty of $0.1(1+z)$.
The signal from lensing alone is shown in the top left panel. The top center panel shows $C^{GI}_{\ell}$ for various different models for intrinsic alignments. We see that the one-halo term becomes important at multipole numbers greater than $\ell\sim 500$
and matches surprisingly well to the NLA result. Including either the one-halo term
or the NLA boosts the GI contamination at $\ell\sim 3000$ by an order of magnitude
relative to that from the linear-alignment model alone.
The II contribution is shown in the top center panel, where we see that the one-halo
term dominates on all scales where the intrinsic alignment contribution is significant. At the smallest scales considered the one-halo term is a few times larger than the NLA model and an order of magnitude larger than the linear alignment model.
However both are much smaller than the GI contribution at small scales.

The total effect is shown in the lower panels of Fig.~\ref{fg:clbin_comparison}.
There is an overall suppression of the power for all the intrinsic alignment
models considered, relative to the alignment-free power spectrum. This is due
to the wide range of redshifts covered by the survey we consider. For a tomographic
survey different contributions would be more important for different pairs of
tomographic bins, as illustrated for the NLA in Ref.~\cite{bridle07}. Overall our halo model result is qualitatively similar to that from the NLA. The suppression is roughly
a factor of two larger at the smallest scales considered. On intermediate scales
the GI suppression is outweighed by the boost in power from II, which is more dominant
at larger scales than GI. The total power spectrum including only the one-halo intrinsic alignment term is actually fractionally larger than the no IA power spectrum at intermediate scales $\ell\sim200$. Then at larger multipole numbers $\ell > 500$ the
GI term dominates again causing a net reduction in power.

\subsection{Bias on $\sigma_8$} 
\label{sub:bias_on_sigma_8_}

\begin{figure*}
\centering
\includegraphics[width=8cm]{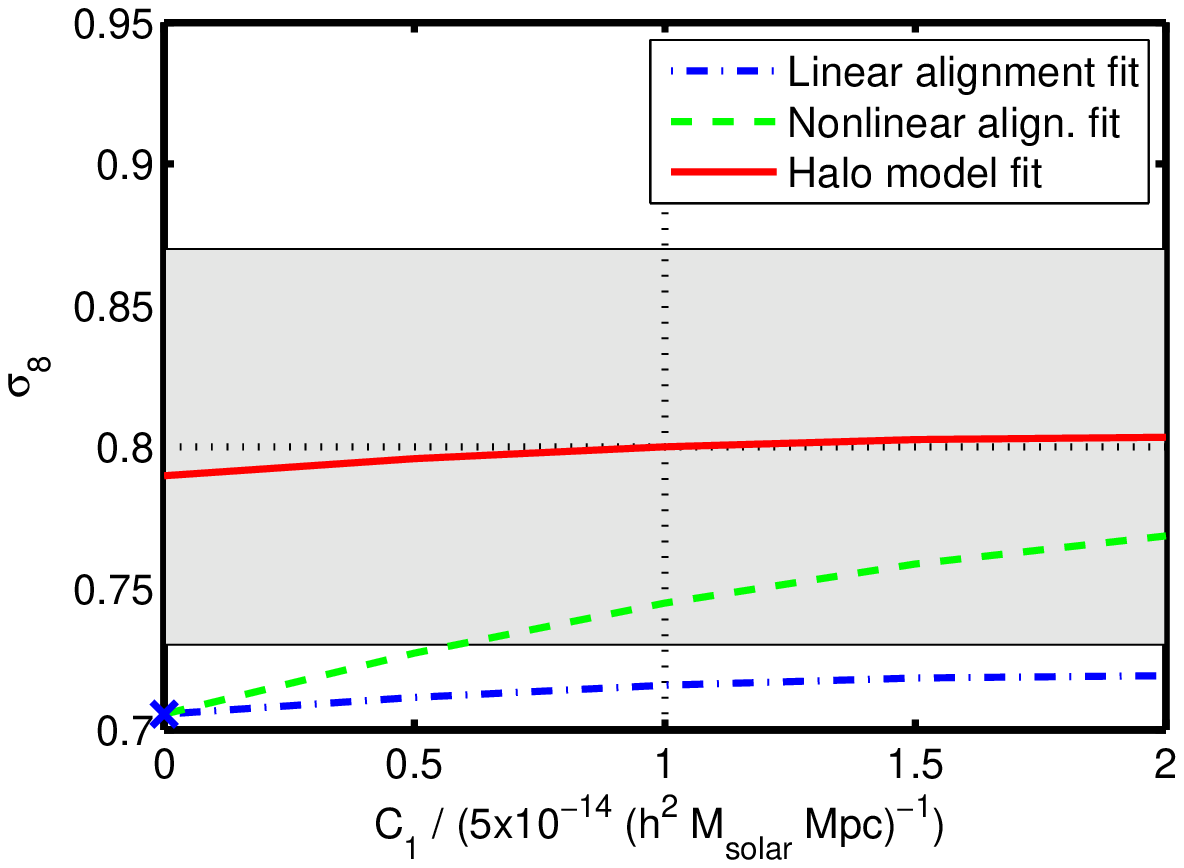}
\includegraphics[width=8cm]{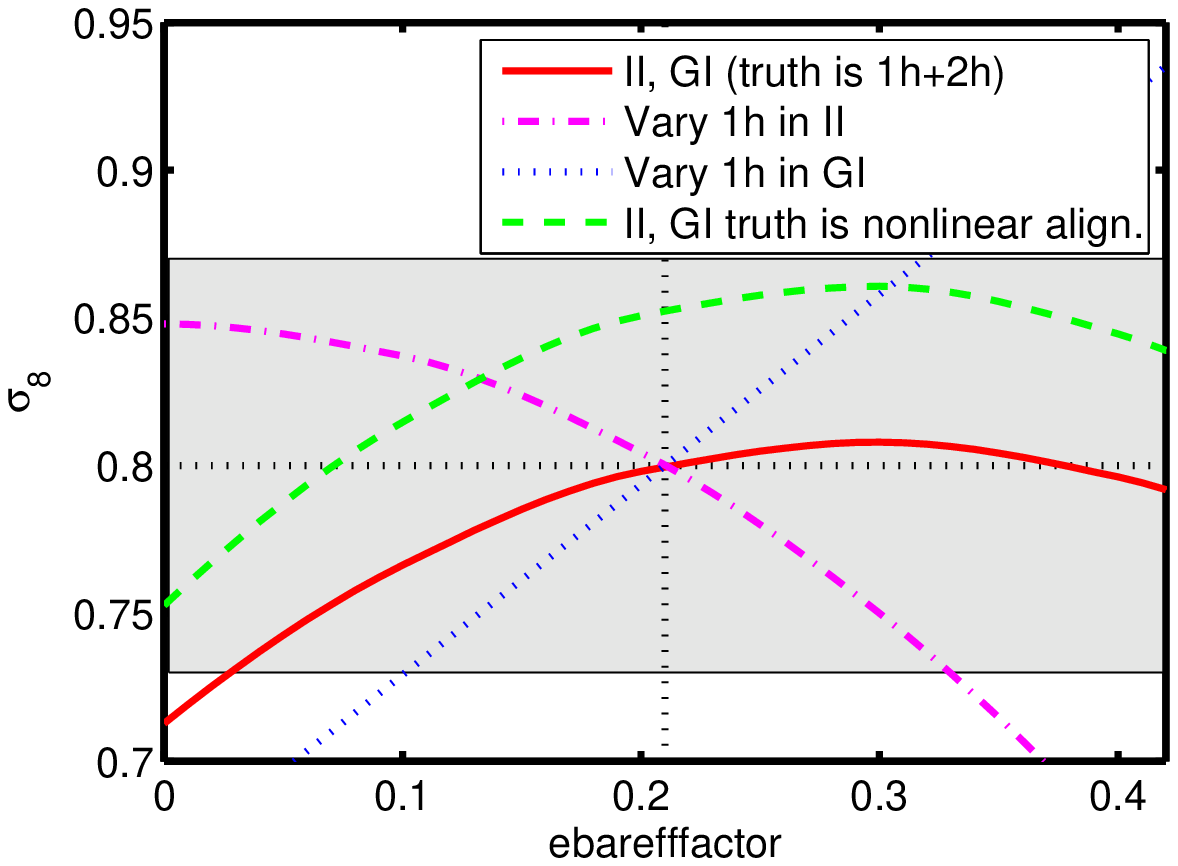}
\caption{\label{fg:parambias}
Matter power spectrum normalisation parameter $\sigma_8$ values that would be inferred when assuming the wrong model for the intrinsic alignment power spectra, as a function of $\cla$ (left) and \eeff (right).  The intersection of the black vertical and horizontal lines shows the fiducial model with $\sigma_8=0.8$, $\cla=5\times 10^{-14} ~\left(h^2 M_{\odot}\,\text{Mpc}^3\right)^{-1}$ and \eeff $=0.21$.
 The solid (red) line shows the inferred $\sigma_8$ values using this model with different assumed $\cla$  or \eeff.
 The dashed line in the left-hand panel shows the result when the non-linear alignment model is fitted to the simulated halo model power spectra.
 The dashed line in the right-hand panel shows the result of fitting the halo model to a nonlinear alignment model set of simulated data.
 The dot-dashed line in the left-hand panel shows the result of fitting the linear alignment model to simulated halo model data. The cross denotes the $\sigma_8$ value that would be obtained on ignoring intrinsic alignments in a naiive fit to data including intrinsic alignments from this halo model (i.e. $C_1=0$ in the linear alignment model).
 In the right hand panel the dotted (blue) and dot-dashed (magenta) lines show the inferred $\sigma_8$ when the theoretical model varies \eeff in only the GI or II terms.
The gray band denotes the rough current uncertainty on $\sigma_8$ from current data, of $\pm 0.07$. }
\end{figure*}

Cosmological parameter constraints from cosmic shear can have systematic biases if our model for the intrinsic alignments contribution to the shear power spectra is wrong or insufficiently flexible.  In this section we illustrate the potential biases on $\sigma_8$
for different assumed intrinsic alignment parameter values.
To find the bias, we calculate the fiducial angular power spectra, the fitted angular power spectra and the Fisher matrix for the fiducial spectra for the assumed survey.
These are combined using Eqn.~(21) of Ref.~\cite{huterertbj06} to find the bias on cosmological parameters. In this case we are varying only one cosmological parameter, $\sigma_8$.
We use a maximum $\ell$ value of $2e4$, 12 galaxies per square arcminute and 100 square degree survey area. 

In the left hand panel of Fig.~(\ref{fg:parambias}) we plot the bias in the inferred $\sigma_8$ value as a function of the amplitude of the linear alignment model, or two-halo term. The fiducial model has our default two-halo term amplitude of $\cla = 5\times 10^{-14} ~\left(h^2 M_{\odot}\,\text{Mpc}^3\right)^{-1}$
and uses the fit functions for $q$ given in the previous section, with $\ebareff=0.21$.
If we incorrectly analysed the data using the linear alignment model we would
bias $\sigma_8$ low by about $0.1$. This is relatively insensitive to the
exact amplitude assumed for the linear alignment model, since the range of
scales probed by the fiducial survey is dominated by the one-halo term (see Fig.~\ref{fg:clbin_comparison}).

An analysis that ignores intrinsic alignments corresponds to assuming
a $\cla=0$ in the linear alignment model, as shown by the leftmost point (cross) of the dashed (or dot-dashed) line in the left hand panel, for which $\sigma_8$ is biased low by just under $0.1$. This is because the fiducial (``observed'') model is contaminated by intrinsic alignments, which are dominated by GI. Therefore the fiducial model power spectrum is lower than it would have been in the absence of intrinsic alignments. Attempting to fit to these ``observed'' points neglecting intrinsic alignments will result in an underestimate of $\sigma_8$ because the lensing-only power spectrum with the true $\sigma_8$ will be too high.

If the NLA model is used when fitting to the data, and yet if the true universe (fiducial model) followed the halo model, then the bias on $\sigma_8$ is reduced because the fitted model contains some of the suppression due to the GI effect. The bias is reduced significantly as the amplitude of the intrinsic alignment model is increased because the NLA mimics to a large extent the shape of the halo model on small scales.
Fig.~\ref{fg:clbin_comparison} shows that the NLA amplitude is generally smaller than that of the halo model, therefore a larger than fiducial $C_1$ is required for the NLA to match the halo model.

On fitting the halo model to the fiducial halo model, the bias on $\sigma_8$ is relatively insensitive to the exact amplitude of the two-halo term. This is for the same reason that the linear alignment model fit is shallow, because the two-halo term has a small weight for this survey.

We have investigated the effect of the assumed amplitude of the
one-halo 
term on biases on $\sigma_8$. In the right hand panel of Fig.~\ref{fg:parambias} we show the impact of assuming the wrong one-halo amplitude, as parameterised by \eeff, the effective 
reduction in 
stick length of the single halo sticks. The full result (solid line) is relatively
complicated so we also show the contributions to this by varying \eeff in only the GI term (dotted line) or in the II term (dot-dashed line).
The dependence of the one-halo GI angular power spectrum amplitude on \eeff is simply linear, and the bias on $\sigma_8$ is roughly linear in the intrinsic alignment perturbation, for small intrinsic alignment contributions.
The dashed line shows that as the amplitude of the GI term is increased in the fitted model then the fitted $\sigma_8$ increases due to the over suppression of the fitted model.

The dependence of the one-halo II angular power spectrum on \eeff is quadratic ($C^{II}_{\ell} \propto $\eeff$^2$) therefore the bias on $\sigma_8$ is also quadratic in \eeff. The direction of the bias is opposite to the GI term alone, because fitting a model with too much II contribution will wrongly boost the total angular power spectrum, and $\sigma_8$ will have to be reduced in the fit to compensate.

Taking the two effects together explains the full result shown by the solid line in the right hand panel of Fig.~\ref{fg:parambias}, in which \eeff is varied self-consistently in both the GI and II terms. The net effect is that an 
underestimate 
for \eeff tends to cause an underestimate of $\sigma_8$.

Finally we consider the case where the fiducial model is the NLA, and the fitted model is the halo model (dashed line). The bias is the same shape as that when the fiducial model is the halo model, but there is an offset (of the same size as the
dashed line in the left hand panel at $C_1=5\times 10^{-14} ~\left(h^2 M_{\odot}\,\text{Mpc}^3\right)^{-1}$). 
We see that the bias is removed on using around half the effective stick length of the fiducial halo model. 

In both panels of  Fig.~\ref{fg:parambias} we compare the biases to the one-sigma $\sigma_8$ error bar of 0.07 obtained by Ref.~\cite{benjamin07}. The largest biases are comparable to this error bar. However, we note that the range of scales used in the actual cosmological parameter analysis 
often 
smaller than the full range used in the angular power spectrum Fisher analysis we use here, and therefore the actual biases from intrinsic alignments will tend to be smaller.



\section{Conclusions} 
\label{sec:conclusions}

We have presented a new model for intrinsic alignments of galaxies. This is inspired by the halo model for the matter and galaxy distribution in the universe and by simulations that suggest that satellite galaxies point towards the centers of halos. We assume the universe is entirely made up of halos that contain (i) a single central galaxy with an orientation determined by the curvature of the large scale potential (the linear alignment model) and (ii) satellite galaxies distributed spherically according to the halo dark matter profile that are oriented pointing at the center of the halo.

We have described several terms that contribute to the intrinsic-intrinsic (II) and lensing-intrinsic (GI) source power spectra. The two-halo term describes contributions to the power spectra from pairs of galaxies in two different halos and can in general contain contributions from central-satellite correlations, satellite-satellite correlations and central-central correlations. We find that the dominant contribution comes from the central-central correlation, and this corresponds to the existing well known linear-alignment model.

The one-halo term describes correlations arising from pairs of galaxies situated in the same halo. In our simple model the central-satellite term is zero and the one-halo term is equal to the satellite-satellite contribution. This is the major result of this paper. The correlations between orientations of galaxies within the same parent halo cause a boost in intrinsic alignment power spectra on small scales. Small scales are not expected to be at all well described by the linear alignment model.

We have considered the impact of random perturbations to the satellite galaxy alignments and found that it reduces the GI power spectrum by around a factor of $0.2$ and the II power spectrum by a factor of $0.2^2$.
We have demonstrated an analytical method for calculating these power spectra based on a multipole expansion. We have also provided fitting formulae for the one-halo term containing 9 parameters for each of the II and GI power spectra.

The intrinsic alignment contributions to cosmic shear angular power spectra are illustrated for a deep tomographic survey and a medium-deep survey using a single redshift bin. As expected the biggest contributions come from the lowest redshifts and for the non-tomographic survey the largest effect is a suppression of power due to the GI alignments.

We made a rough estimate of the possible impact of incorrectly using the wrong intrinsic alignment model for cosmological parameter constraints from a medium-deep survey. We find that, if intrinsic alignments are ignored completely in the cosmological parameter analysis, and all available scales are included in the analysis, then estimates of the matter power spectrum normalisation $\sigma_8$ can be biased low by approximately the current 1-$\sigma$ uncertainty.
However, we note that cosmological parameter constraints are usually performed using a limited range of scales, and therefore will be much less sensitive to the small scale effects introduced in this paper and $\sigma_8$ will be much less biased. Investigation of how the bias depends on the exact analysis is beyond the scope of this work.

Our implementation of the halo model we present is subject to a number of limitations.
We assume all halos are spherical and all satellites point at the halo center, subject to a random misalignment angle. Future work could investigate the impact of using elliptical halos in which
satellites are anisotropically distributed and
the satellite orientations depend on their position within the halo~\cite{faltenbacher07a}.
We assume there is only a single population of satellite galaxies, and a single population of central galaxies. However a more sophisticated approach could use a mixture of spiral and elliptical galaxies with different alignment models for each.
We have also assumed that the ellipticity of the satellite galaxies is independent of radius and independent of the mass of the parent halo~\cite{pereira08}.
Finally we have assumed that the central galaxies are aligned with their host halo, whereas in fact there could be additional randomness~\cite{faltenbacher09}.

The halo model implementation could be improved considerably in the future as more fitting functions become available from simulators.
Due to this separation into one and two-halo terms, significant progress can be made using high resolution simulations of single parent halos instead of full-sized cosmological simulations.
The fitting formula for the distribution of galaxy alignment angles from Ref.~\cite{knebe08} was very useful.
It would be helpful to in addition implement fitting formulae for
(i) how satellite galaxy ellipticity depends on radius within the parent halo, and on
parent halo mass and
(ii) how the distribution in alignment angles changes with radius and parent halo mass.
For future extentions to this halo model it would also be useful to have
(iii) the distribution in alignment angles between galaxies and dark matter halos (when using N-body simulation results to construct a model for galaxies), 
(iv) fitting formualae for how the satellite galaxy orientation depends on position within the halo for non-spherical halos,
(v) the alignment distribution between satellites and the central galaxy in a halo, and
(vi) a dependence of all the above on satellite mass and/or type. 
Once point (v) is known, a central-satellite one-halo term could potentially be added to our results using the distribution between the ellipticity of the central region of a halo and that of the total halo recently given in Ref.~\cite{faltenbacher09}.  
To build a halo model for central galaxies that could replace the LA model used here for the two-halo term, one would also need to know the IA correlation function between elliptical halos.  We will pursue this in future work.


\appendix

\section{Normalization when sources are clustered} 
\label{sec:marked_power_spectra}

Because galaxy positions are correlated, we must normalize the ellipticity power spectra from Section~\ref{sec:intrinsic_ellipticity_power_spectra} by the number of galaxy pairs at a given angular separation,
\[
N_{\text{pairs}}(\theta) = \nbar^2_{\text{gal}}\left(1+\omega(\theta)\right),
\]
where $\omega(\theta)$ is the galaxy angular correlation function.  Applying this normalization to the ellipticity correlation function, we get,
\begin{equation}\label{eq:normxi}
\xi_{\pm}(\theta,\chi) = \int_{0}^{\infty} \frac{\ell d\ell}{2\pi}\,
\frac{J_{0.4}(\ell\theta)}{1+\omega(\theta)}
\left[P^{EE}_{\densshear}(\ell,\chi)
\pm P^{BB}_{\densshear}(\ell,\chi) \right].
\end{equation}
Following a similar derivation in Ref.~\cite{schneider02}, we can rearrange the multiplicative normalization into an additive correction using,
\begin{equation}
\frac{1}{1+\omega(\theta)} = 1 - \frac{\omega(\theta)}{1+\omega(\theta)}.
\end{equation}
To get the additive correction to the ellipticity power spectra,
we put Eqn.~(\ref{eq:normxi}) into the transformation,
\begin{equation}
P^{EE,BB}_{\densshear}
(\ell) = \pi\int_{0}^{\infty} \theta d\theta\, \left[
\xi_{+}(\theta) J_{0}(\ell\theta) \pm \xi_{-}(\theta) J_{4}(\ell\theta)
\right],
\end{equation}
to get the modification to the power spectra due to the extra $\omega(\theta)/(1+\omega(\theta)$,
\begin{multline}
\Delta \Pgi^{EE,BB}
(\ell,\fA) = -\half\int_{0}^{\infty} \ell'\,d\ell'\,
\{ J^{I0}(\ell,\ell')
\left[P^{EE}(\ell')
\right.\\ \quad \left.
+P^{BB}(\ell')
\right]
\pm
J^{I4}(\ell,\ell')\left[
P^{EE}(\ell')-P^{BB}
(\ell')\right] \},
\end{multline}
with
\begin{equation}
\Delta P_{\delta,\densshear}(\ell,\fA) = 2\pi\int_{0}^{\infty} \ell'\,d\ell'\,
P_{\delta,\densshear}(\ell',\fA)\, J^{I2}(\ell,\ell'),
\end{equation}
and
\begin{equation}\label{eq:prodbessel}
J^{I\nu}(\ell,\ell') \equiv \int_{0}^{\infty} d\theta\, \theta\,
\frac{\omega(\theta)}{1+\omega(\theta)}\, J_{\nu}(\ell\theta)
J_{\nu}(\ell'\theta).
\end{equation}

Again following Ref.~\cite{schneider02}, if we assume $\omega(\theta)\ll 1$, $\omega(\theta)/(1+\omega(\theta))\approx \omega(\theta)$.  If we further approximate $\omega(\theta)$ as a power law in $\theta$, then the integral in Eqn.~(\ref{eq:prodbessel}) can be done with Eqns.~11.4.33, 11.4.34 in Ref.~\cite{abromowitz+stegun}.  We use the (slightly modified) power-law model from Eqn.~(66) of Ref.~\cite{schneider02},
\begin{equation}
\omega(\theta) \approx A \left(\frac{\theta}{\text{1 arcmin.}}\right)^{1-\gamma}
\end{equation}
with $\gamma=1.7$ and $A=0.5$.

With this model, we find negligible corrections to the unnormalized spectra given in Section~\ref{sec:intrinsic_ellipticity_power_spectra} 
and thus we neglect the effect of $\omega(\theta)$ in Eqn.~(\ref{eq:normxi}).

\section{Computation of ellipticity power spectra} 
\label{sec:computation_of_ellipticity_power_spectra}
In this section we first derive the exact dependence of the satellite ellipticity model on the radial alignment angles $\beta$ and $\eta$ from Eqns.~(\ref{eq:Knebedist}) and (\ref{eq:etadist}) and then show how we compute the power spectra by means of a multipole expansion of the satellite ellitpicity distribution in a halo.

\subsection{Galaxy major axis as a function of radial alignment angles} 
\label{sub:galaxy_major_axis_as_a_function_of_radial_alignment_angles}
First consider the explicit dependence of the ellipticity of the satellite galaxies on the alignment angles $\beta$ and $\eta$ from Section~\ref{sub:distribution_in_radial_alignment_angles}.
If we write
\begin{equation}\label{eq:stickaxiscoords}
\hat{e}=\left(\sin\theta_e \cos\phi_e, \sin\theta_e \sin\phi_e, \cos\theta_e\right),
\end{equation}
then the ellipticity of a satellite galaxy in a halo is,
\begin{equation}
\gamma(\rv) = \ebar(r,m) \sin\theta_e \left(\cos2\phi_e,\sin2\phi_e\right) .
\end{equation}
To find $\theta_e$ and $\phi_e$ as functions of $\theta$, $\phi$, $\beta$, and $\eta$, we use a set of Euler angles to rotate to a coordinate system $\rv'$ with $\hat{z}'=\hat{r}$. This can be accomplished by setting the Euler angles (in the $zxz$ convention) to $\alpha^E=\phi+\pi/2$, $\beta^E=\theta$, $\gamma^E=-\pi/2$.
\begin{widetext}
Explicitly, the components of the unit vector describing the major axis of a galaxy in the 2 coordinate systems are related by
\begin{align}
\hat{e}' &= \left(\sin\beta\cos\eta,\sin\beta\sin\eta,\cos\beta\right)^T
\notag\\
&= R(\alpha^E,\beta^E,\gamma^E)\,\hat{e}
\notag\\
&\equiv \left(
\begin{array}{ccc}
  \cos\gamma^E & -\sin\gamma^E & 0 \\
  \sin\gamma^E & \cos\gamma^E & 0 \\
  0 & 0 & 1
\end{array}\right)
\left(
\begin{array}{ccc}
  1 & 0 & 0 \\
  0 & \cos\beta^E & -\sin\beta^E \\
  0 & \sin\beta^E & \cos\beta^E
\end{array}\right)
\left(
\begin{array}{ccc}
  \cos\alpha^E & -\sin\alpha^E & 0 \\
  \sin\alpha^E & \cos\alpha^E & 0 \\
  0 & 0 & 1
\end{array}\right) \,
\hat{e}.
\end{align}
\end{widetext}
We then have $\hat{e} = R^{-1}\hat{e}'$ and
\begin{equation}\label{eq:thetae_phie}
\sin\theta_e = \sqrt{\hat{e}_1^2 + \hat{e}_2^2},\quad
\cos\phi_e = \frac{\hat{e}_1}{\sin\theta_e},\quad
\sin\phi_e = \frac{\hat{e}_2}{\sin\theta_e},
\end{equation}
giving $\sin\theta_e$, $\cos\phi_e$, and $\sin\phi_e$ as functions of $\theta$, $\phi$, $\beta$, and $\eta$ as desired.

\begin{figure*}
  \centering
  \includegraphics[height=5.7cm]{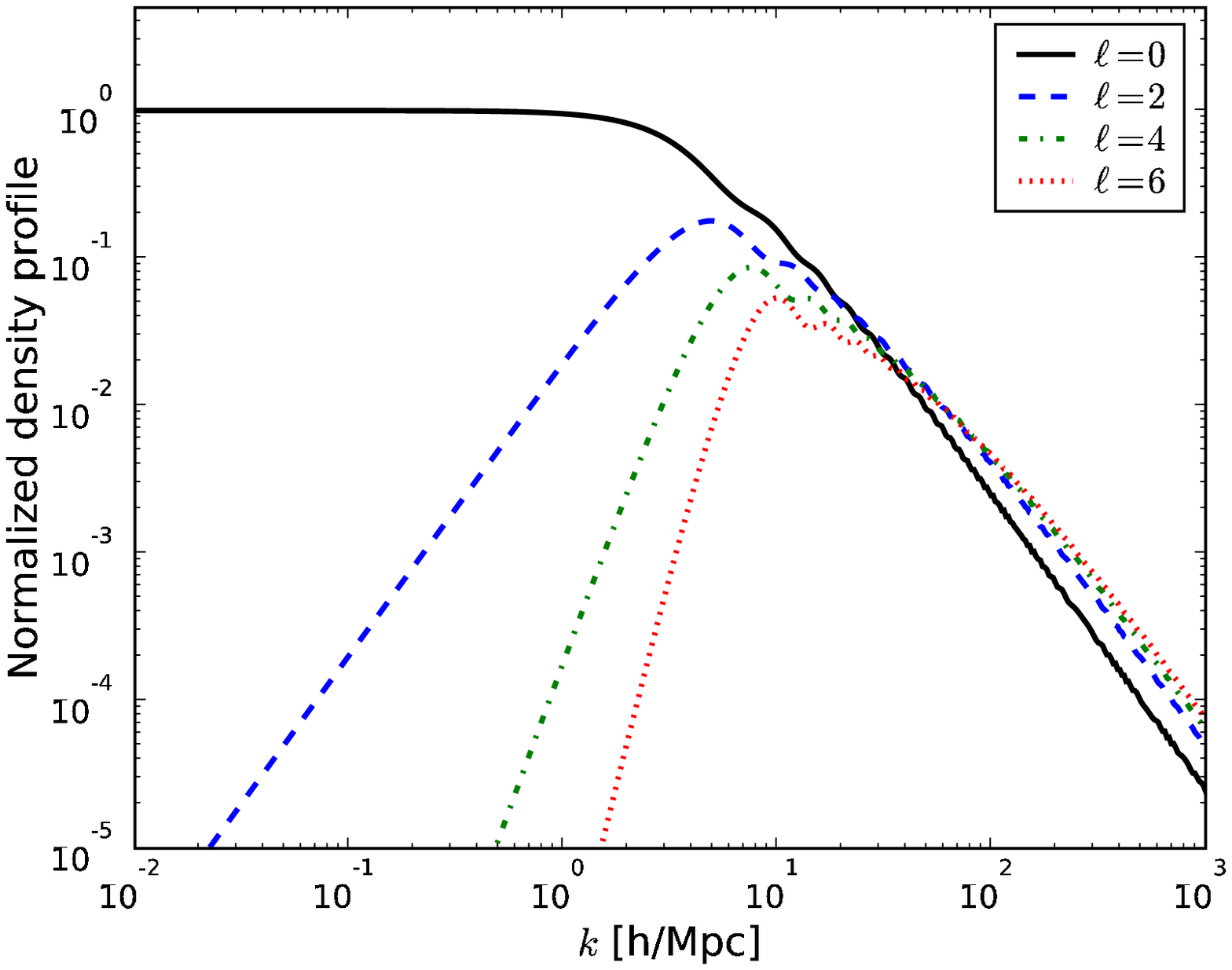}
  \includegraphics[height=6cm]{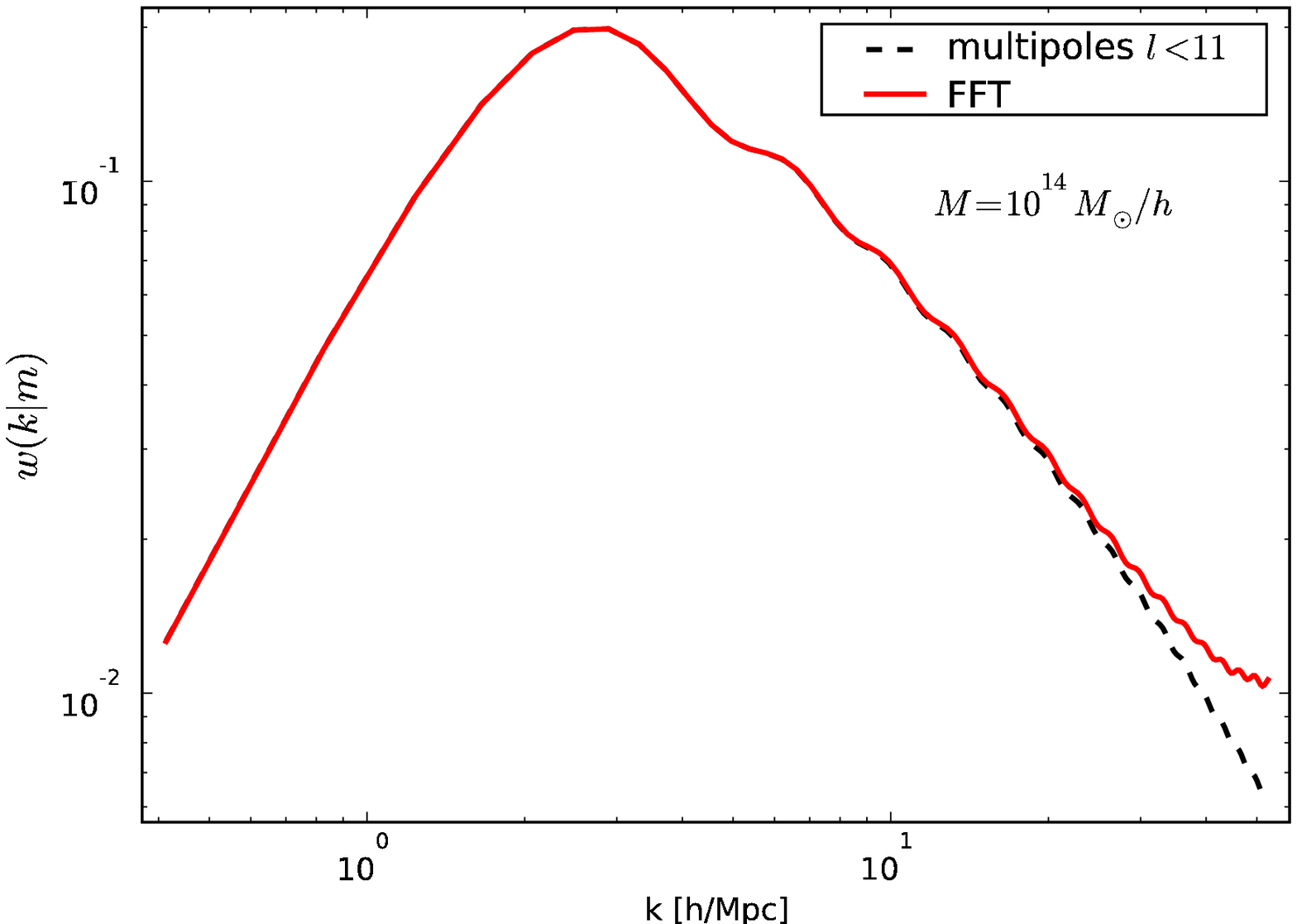}
  \caption{\label{fg:NFWmultipoleTests}
  The left hand panel shows the first 4 nonzero multipole moments of the normalized NFW profile.  Higher multipoles are subdominant for small wavenumber.
  The right hand panel shows a comparison of the Fourier transform of $\densshear$ for the stick model for satellite galaxies with perfect radial alignments computed using a 3-D FFT (red, solid) and the $\ell=2,4,6,8,10$ multipoles of the NFW profile.}
\end{figure*}

\subsection{Satellite ellipticity density run} 
\label{sub:satellite_ellipticity_density_run}
We define the normalized density run of the satellite ellipticity distribution in a halo as
\begin{equation}\label{eq:densRunDefinition}
w(\kv|m) \equiv \frac{\int d^{3}r\,\intshear(\rv,m)\,\rhonfw(r,m,c)\,e^{i\kv\cdot\rv}}{\int d^{3}r\,\left|\intshear(\rv,m)\right|\,\rhonfw(r,m,c)},
\end{equation}
where, when allowing for a distribution in radial alignments,
\begin{equation}
\intshear(\rv,m) = \ebar(m)\,\sin\left(\theta_e(\theta,\phi,\beta,\eta)\right)
\exp\left[2i\phi_e\left(\theta,\phi,\beta,\eta\right) \right],
\end{equation}
where $\theta_e$ and $\phi_e$ describe the orientation of the galaxy major axis as in Eqn.~(\ref{eq:stickaxiscoords}), and we have assumed that the magnitude of the intrinsic ellipticity, $\ebar(m)$, is independent of position within the halo.
Separating the two components of the complex ellipticity,
\begin{align}
w_{1}(\kv|m) &= \frac{4}{\pi}\int d^{3}r\,\costwophie\,\sin(\theta_e) u(r|m) e^{i\kv\cdot\rv},
\notag\\
w_{2}(\kv|m) &= \frac{4}{\pi}\int d^{3}r\,\sintwophie\,\sin(\theta_e) u(r|m) e^{i\kv\cdot\rv}.
\end{align}

We can simplify the computation of this transform by performing a multipole expansion,
\begin{align}\label{eq:multipoleexpansion}
   w(\kv|m) &= \sum_{\ell=0}^{\infty} (2\ell+1)\,i^{\ell}\,u_{\ell}(k|m)
   \notag\\
   &\times\left[ \int_{-1}^{1}d\cos\theta\,
   \int_{0}^{2\pi}d\phi\, \sin(\theta_e)\,e^{2i\phi_e} P_{\ell}(\cos\gamma) \right]
   \nonumber\\
   &\equiv \sum_{\ell=0}^{\infty} (2\ell+1)\,i^{\ell}\,u_{\ell}(k|m)
   f_{\ell}(\theta_{k},\phi_{k},\beta,\eta)
\end{align}
where $\cos\gamma = \sin\theta_{k}\,\sin\theta\,\cos(\phi_{k}-\phi) + \cos\theta_{k}\,\cos\theta$ and
\[
  u_{\ell}(k|m) \equiv \int_{0}^{\infty} r^{2}dr\, u(r|m)\, j_{\ell}(kr)
\]
are normalized multipole moments of the NFW profile.  The integral over $\phi$ in $f_{\ell}$ guarantees that $f_{\ell}=0$ for odd $\ell$ (and $\ell=0$).  The first four even multipoles $u_{\ell}(k|m)$ are shown in the left hand panel of Fig.\ref{fg:NFWmultipoleTests}.
If we assume that the radial mis-alignment of the galaxies is uncorrelated such that for 2 galaxies $P(\beta_1,\eta_1,\beta_2,\eta_2) = P(\beta_1,\eta_1)\,P(\beta_2,\eta_2)$, then this expression for $f_{\ell}$ can be integrated over $\beta$ and $\eta$ at this stage before being used to compute correlation functions.

For the case of $\beta=0$ (so $\theta_e=\theta$ and $\phi_e=\phi$), all the integrals over $\phi$ in $f_{\ell}$ are of the form,
\begin{equation}
   \int_{0}^{2\pi}d\phi\,
   \left(\begin{array}{c}
   \cos 2\phi\\
   \sin 2\phi
   \end{array}\right)
   \cos^{n} \left(\phik-\phi\right)
   \equiv
   \left(\begin{array}{c}
   \cos 2\phik\\
   \sin 2\phik
   \end{array}\right)
   \,g_{n}
\end{equation}
with, for example,
\begin{equation}
   g_2 = g_4 = \frac{\pi}{2},\,
   g_6 = \frac{15\pi}{32},\,
   g_8 = \frac{7\pi}{16},\,
   g_{10} = \frac{105\pi}{256}.
\end{equation}
This makes it clear that the only $\phik$ dependence in $f_{\ell}$ is in the $\cos$/$\sin$ phase factors.

We can then use Eqn.~(\ref{eq:EBmodes}) to write,
\begin{align}
  \densshear_{E}(\kv) &= \left(\cos^2(2\phi_k)
  + \sin^2(2\phi_k)\right)\,\left|\densshear(\kv)\right|
  \notag\\
  &= \left|\densshear(\kv)\right|
  \notag\\
  \densshear_{B}(\kv) &= \left(\cos(2\phi_k)\sin(2\phi_k)
  -\cos(2\phi_k)\sin(2\phi_k)\right)\,\left|\densshear(\kv)\right|
  \notag\\
  &= 0.
\end{align}
So, the stick model with perfect radial alignments for satellite galaxies has zero B-mode.

Again for $\beta=0$, the angular term in the multipole expansion, $f_{\ell}$, can be computed using the formula,
\begin{align}
  f_{\ell}(\theta_k,\phi_k) &= e^{2i\phi_k} \sum_{m=0}^{\ell} p_{lm}\,
  \sum_{j=0}^{\ell-m}
  \left(\begin{array}{c}\ell-m \\ j \end{array}\right)
  g_j\,
  \notag\\
  &\times\sin^{j}(\theta_k)\,\cos^{\ell-m-j}(\theta_k)\,
  I(j+1,\ell-m-j)
  \end{align}
  where $p_{lm}$ are coefficients of the $\ell$th Legendre polynomial and
  \begin{equation}
  I(a,b) \equiv \int_{-1}^{1}dx\, \left(1-x^2\right)^{\frac{a}{2}}x^b.
\end{equation}

We compare Eqn.~(\ref{eq:multipoleexpansion}) for $\ell\le10$ with the exact computation of Eqn.~(\ref{eq:densRunDefinition}) in
the right hand panel of Fig.~\ref{fg:NFWmultipoleTests}.  We find that the sum over multipoles converges rapidly on the scales of interest. All of the results in the main body of the paper use only the $\ell=2$ term when computing $w(\kv|m)$.



\begin{acknowledgments}
We thank the organizers and participants of the workshop on ``Intrinsic Alignments and Cosmic Shear'' at UCL 31 March to 4 April 2008 for many ideas and discussions that led to the inception of this work.  We would also like to thank Robert Smith for pointing us to his work on modeling elliptical halo profiles.
 We thank Rachel Mandelbaum, David Spergel, Adam Amara and Oliver Hahn for helpful conversations. 
 SLB acknowledges the Royal Society for support in the form of a University Research Fellowship, and CEA Saclay for hospitality while part of this work was carried out.
 MDS thanks the British Council for the Researchers Exchange Programme award that initiated this collaboration.
\end{acknowledgments}

\def\urlprefix{}
\def\url#1{}
\bibliographystyle{apsrev}
\bibliography{iaHOD}

\end{document}

%% file: Pkfitparamsfit.tex
\begin{table}[hb]
\centering
\caption{\label{tb:PkIIfitparamsfit}Fitting function parameters for $p_{i}$ in the II power spectrum fit.}
\begin{ruledtabular}
\begin{tabular}{cccc}
param. index & $q_{i1}$ & $q_{i2}$ & $q_{i3}$\\
\hline
1 & 0.09939 & 3.718 & 0.3475\\
2 & 1.931 & 1.061 & 0.7484\\
3 & 6.082 & 0.1045 & 0.613\\
\end{tabular}
\end{ruledtabular}
\end{table}

%% file: PkGIfitparamsfit.tex
\begin{table}[hb]
\centering
\caption{\label{tb:PkGIfitparamsfit}Fitting function parameters for $p_{i}$ in the GI power spectrum fit.}
\begin{ruledtabular}
\begin{tabular}{cccc}
param. index & $q_{i1}$ & $q_{i2}$ & $q_{i3}$\\
\hline
1 & 0.01867 & 6.924 & 0.3725\\
2 & 1.989 & 1.081 & 0.6816\\
3 & 4.232 & 0.1748 & 0.481\\
\end{tabular}
\end{ruledtabular}
\end{table}

%% file: ms.bbl
\begin{thebibliography}{52}
\expandafter\ifx\csname natexlab\endcsname\relax\def\natexlab#1{#1}\fi
\expandafter\ifx\csname bibnamefont\endcsname\relax
  \def\bibnamefont#1{#1}\fi
\expandafter\ifx\csname bibfnamefont\endcsname\relax
  \def\bibfnamefont#1{#1}\fi
\expandafter\ifx\csname citenamefont\endcsname\relax
  \def\citenamefont#1{#1}\fi
\expandafter\ifx\csname url\endcsname\relax
  \def\url#1{\texttt{#1}}\fi
\expandafter\ifx\csname urlprefix\endcsname\relax\def\urlprefix{URL }\fi
\providecommand{\bibinfo}[2]{#2}
\providecommand{\eprint}[2][]{\url{#2}}

\bibitem[{\citenamefont{{Albrecht} et~al.}(2006)\citenamefont{{Albrecht},
  {Bernstein}, {Cahn}, {Freedman}, {Hewitt}, {Hu}, {Huth}, {Kamionkowski},
  {Kolb}, {Knox} et~al.}}]{detf}
\bibinfo{author}{\bibfnamefont{A.}~\bibnamefont{{Albrecht}}},
  \bibinfo{author}{\bibfnamefont{G.}~\bibnamefont{{Bernstein}}},
  \bibinfo{author}{\bibfnamefont{R.}~\bibnamefont{{Cahn}}},
  \bibinfo{author}{\bibfnamefont{W.~L.} \bibnamefont{{Freedman}}},
  \bibinfo{author}{\bibfnamefont{J.}~\bibnamefont{{Hewitt}}},
  \bibinfo{author}{\bibfnamefont{W.}~\bibnamefont{{Hu}}},
  \bibinfo{author}{\bibfnamefont{J.}~\bibnamefont{{Huth}}},
  \bibinfo{author}{\bibfnamefont{M.}~\bibnamefont{{Kamionkowski}}},
  \bibinfo{author}{\bibfnamefont{E.~W.} \bibnamefont{{Kolb}}},
  \bibinfo{author}{\bibfnamefont{L.}~\bibnamefont{{Knox}}},
  \bibnamefont{et~al.}, \bibinfo{journal}{ArXiv Astrophysics e-prints}
  (\bibinfo{year}{2006}), \eprint{astro-ph/0609591}.

\bibitem[{\citenamefont{{Peacock} and {Schneider}}(2006)}]{esoesa}
\bibinfo{author}{\bibfnamefont{J.}~\bibnamefont{{Peacock}}} \bibnamefont{and}
  \bibinfo{author}{\bibfnamefont{P.}~\bibnamefont{{Schneider}}},
  \bibinfo{journal}{The Messenger} \textbf{\bibinfo{volume}{125}},
  \bibinfo{pages}{48} (\bibinfo{year}{2006}).

\bibitem[{\citenamefont{{Heavens} and {Peacock}}(1988)}]{heavensp88}
\bibinfo{author}{\bibfnamefont{A.}~\bibnamefont{{Heavens}}} \bibnamefont{and}
  \bibinfo{author}{\bibfnamefont{J.}~\bibnamefont{{Peacock}}},
  \bibinfo{journal}{\mnras} \textbf{\bibinfo{volume}{232}},
  \bibinfo{pages}{339} (\bibinfo{year}{1988}).

\bibitem[{\citenamefont{{Schaefer}}(2008)}]{schaefer08}
\bibinfo{author}{\bibfnamefont{B.~M.} \bibnamefont{{Schaefer}}}
  (\bibinfo{year}{2008}), \eprint{arXiv:0808.0203}.

\bibitem[{\citenamefont{{Catelan}
  et~al.}(2001{\natexlab{a}})\citenamefont{{Catelan}, {Kamionkowski}, and
  {Blandford}}}]{catelankb01}
\bibinfo{author}{\bibfnamefont{P.}~\bibnamefont{{Catelan}}},
  \bibinfo{author}{\bibfnamefont{M.}~\bibnamefont{{Kamionkowski}}},
  \bibnamefont{and} \bibinfo{author}{\bibfnamefont{R.~D.}
  \bibnamefont{{Blandford}}}, \bibinfo{journal}{\mnras}
  \textbf{\bibinfo{volume}{320}}, \bibinfo{pages}{L7}
  (\bibinfo{year}{2001}{\natexlab{a}}), \eprint{astro-ph/0005470}.

\bibitem[{\citenamefont{{Aubert} et~al.}(2004)\citenamefont{{Aubert}, {Pichon},
  and {Colombi}}}]{aubert04}
\bibinfo{author}{\bibfnamefont{D.}~\bibnamefont{{Aubert}}},
  \bibinfo{author}{\bibfnamefont{C.}~\bibnamefont{{Pichon}}}, \bibnamefont{and}
  \bibinfo{author}{\bibfnamefont{S.}~\bibnamefont{{Colombi}}},
  \bibinfo{journal}{\mnras} \textbf{\bibinfo{volume}{352}},
  \bibinfo{pages}{376} (\bibinfo{year}{2004}), \eprint{arXiv:astro-ph/0402405}.

\bibitem[{\citenamefont{Crittenden et~al.}(2001)\citenamefont{Crittenden,
  Natarajan, Pen, and Theuns}}]{crittenden01}
\bibinfo{author}{\bibfnamefont{R.~G.} \bibnamefont{Crittenden}},
  \bibinfo{author}{\bibfnamefont{P.}~\bibnamefont{Natarajan}},
  \bibinfo{author}{\bibfnamefont{U.-L.} \bibnamefont{Pen}}, \bibnamefont{and}
  \bibinfo{author}{\bibfnamefont{T.}~\bibnamefont{Theuns}},
  \bibinfo{journal}{The Astrophysical Journal} \textbf{\bibinfo{volume}{559}},
  \bibinfo{pages}{552} (\bibinfo{year}{2001}), \bibinfo{note}{(c) 2001: The
  American Astronomical Society},
  \urlprefix\url{http://adsabs.harvard.edu/cgi-bin/nph-data_query?bibcode=2001%
ApJ...559..552C&link_type=ABSTRACT}.

\bibitem[{\citenamefont{Crittenden et~al.}(2002)\citenamefont{Crittenden,
  Natarajan, Pen, and Theuns}}]{crittenden02}
\bibinfo{author}{\bibfnamefont{R.~G.} \bibnamefont{Crittenden}},
  \bibinfo{author}{\bibfnamefont{P.}~\bibnamefont{Natarajan}},
  \bibinfo{author}{\bibfnamefont{U.-L.} \bibnamefont{Pen}}, \bibnamefont{and}
  \bibinfo{author}{\bibfnamefont{T.}~\bibnamefont{Theuns}},
  \bibinfo{journal}{The Astrophysical Journal} \textbf{\bibinfo{volume}{568}},
  \bibinfo{pages}{20} (\bibinfo{year}{2002}), \bibinfo{note}{(c) 2002: The
  American Astronomical Society},
  \urlprefix\url{http://cdsads.u-strasbg.fr/cgi-bin/nph-data_query?bibcode=200%
2ApJ...568...20C&link_type=ABSTRACT}.

\bibitem[{\citenamefont{Hirata and Seljak}(2004)}]{hirata04}
\bibinfo{author}{\bibfnamefont{C.~M.} \bibnamefont{Hirata}} \bibnamefont{and}
  \bibinfo{author}{\bibfnamefont{U.}~\bibnamefont{Seljak}},
  \bibinfo{journal}{Phys. Rev. D} \textbf{\bibinfo{volume}{70}},
  \bibinfo{pages}{63526} (\bibinfo{year}{2004}),
  \urlprefix\url{http://adsabs.harvard.edu/cgi-bin/nph-data_query?bibcode=2004%
PhRvD..70f3526H&link_type=ABSTRACT}.

\bibitem[{\citenamefont{Heavens et~al.}(2000)\citenamefont{Heavens, Refregier,
  and Heymans}}]{heavens00}
\bibinfo{author}{\bibfnamefont{A.}~\bibnamefont{Heavens}},
  \bibinfo{author}{\bibfnamefont{A.}~\bibnamefont{Refregier}},
  \bibnamefont{and} \bibinfo{author}{\bibfnamefont{C.}~\bibnamefont{Heymans}},
  \bibinfo{journal}{Monthly Notices RAS} \textbf{\bibinfo{volume}{319}},
  \bibinfo{pages}{649} (\bibinfo{year}{2000}), \bibinfo{note}{(c) 2000 The
  Royal Astronomical Society},
  \urlprefix\url{http://adsabs.harvard.edu/cgi-bin/nph-data_query?bibcode=2000%
MNRAS.319..649H&link_type=ABSTRACT}.

\bibitem[{\citenamefont{{Croft} and {Metzler}}(2000)}]{croftm00}
\bibinfo{author}{\bibfnamefont{R.~A.~C.} \bibnamefont{{Croft}}}
  \bibnamefont{and} \bibinfo{author}{\bibfnamefont{C.~A.}
  \bibnamefont{{Metzler}}}, \bibinfo{journal}{\apj}
  \textbf{\bibinfo{volume}{545}}, \bibinfo{pages}{561} (\bibinfo{year}{2000}),
  \eprint{astro-ph/0005384}.

\bibitem[{\citenamefont{{Lee} et~al.}(2008)\citenamefont{{Lee}, {Springel},
  {Pen}, and {Lemson}}}]{leespl08}
\bibinfo{author}{\bibfnamefont{J.}~\bibnamefont{{Lee}}},
  \bibinfo{author}{\bibfnamefont{V.}~\bibnamefont{{Springel}}},
  \bibinfo{author}{\bibfnamefont{U.-L.} \bibnamefont{{Pen}}}, \bibnamefont{and}
  \bibinfo{author}{\bibfnamefont{G.}~\bibnamefont{{Lemson}}},
  \bibinfo{journal}{\mnras} \textbf{\bibinfo{volume}{389}},
  \bibinfo{pages}{1266} (\bibinfo{year}{2008}), \eprint{0709.1106}.

\bibitem[{\citenamefont{{Heymans} et~al.}(2006)\citenamefont{{Heymans},
  {White}, {Heavens}, {Vale}, and {van Waerbeke}}}]{heymans06}
\bibinfo{author}{\bibfnamefont{C.}~\bibnamefont{{Heymans}}},
  \bibinfo{author}{\bibfnamefont{M.}~\bibnamefont{{White}}},
  \bibinfo{author}{\bibfnamefont{A.}~\bibnamefont{{Heavens}}},
  \bibinfo{author}{\bibfnamefont{C.}~\bibnamefont{{Vale}}}, \bibnamefont{and}
  \bibinfo{author}{\bibfnamefont{L.}~\bibnamefont{{van Waerbeke}}},
  \bibinfo{journal}{\mnras} \textbf{\bibinfo{volume}{371}},
  \bibinfo{pages}{750} (\bibinfo{year}{2006}), \eprint{arXiv:astro-ph/0604001}.

\bibitem[{\citenamefont{{Hawley} and {Peebles}}(1975)}]{hawley75}
\bibinfo{author}{\bibfnamefont{D.~L.} \bibnamefont{{Hawley}}} \bibnamefont{and}
  \bibinfo{author}{\bibfnamefont{P.~J.~E.} \bibnamefont{{Peebles}}},
  \bibinfo{journal}{\aj} \textbf{\bibinfo{volume}{80}} (\bibinfo{year}{1975}).

\bibitem[{\citenamefont{{Ciotti} and {Dutta}}(1994)}]{ciotti94}
\bibinfo{author}{\bibfnamefont{L.}~\bibnamefont{{Ciotti}}} \bibnamefont{and}
  \bibinfo{author}{\bibfnamefont{S.~N.} \bibnamefont{{Dutta}}},
  \bibinfo{journal}{\mnras} \textbf{\bibinfo{volume}{270}},
  \bibinfo{pages}{390} (\bibinfo{year}{1994}), \eprint{arXiv:astro-ph/9404059}.

\bibitem[{\citenamefont{{Pereira} and {Kuhn}}(2005)}]{pereira05}
\bibinfo{author}{\bibfnamefont{M.~J.} \bibnamefont{{Pereira}}}
  \bibnamefont{and} \bibinfo{author}{\bibfnamefont{J.~R.}
  \bibnamefont{{Kuhn}}}, \bibinfo{journal}{\apjl}
  \textbf{\bibinfo{volume}{627}}, \bibinfo{pages}{L21} (\bibinfo{year}{2005}),
  \eprint{arXiv:astro-ph/0411710}.

\bibitem[{\citenamefont{{Agustsson} and {Brainerd}}(2006)}]{agustsson06}
\bibinfo{author}{\bibfnamefont{I.}~\bibnamefont{{Agustsson}}} \bibnamefont{and}
  \bibinfo{author}{\bibfnamefont{T.~G.} \bibnamefont{{Brainerd}}},
  \bibinfo{journal}{\apjl} \textbf{\bibinfo{volume}{644}}, \bibinfo{pages}{L25}
  (\bibinfo{year}{2006}), \eprint{arXiv:astro-ph/0509405}.

\bibitem[{\citenamefont{{Faltenbacher}
  et~al.}(2007)\citenamefont{{Faltenbacher}, {Li}, {Mao}, {van den Bosch},
  {Yang}, {Jing}, {Pasquali}, and {Mo}}}]{faltenbacher07a}
\bibinfo{author}{\bibfnamefont{A.}~\bibnamefont{{Faltenbacher}}},
  \bibinfo{author}{\bibfnamefont{C.}~\bibnamefont{{Li}}},
  \bibinfo{author}{\bibfnamefont{S.}~\bibnamefont{{Mao}}},
  \bibinfo{author}{\bibfnamefont{F.~C.} \bibnamefont{{van den Bosch}}},
  \bibinfo{author}{\bibfnamefont{X.}~\bibnamefont{{Yang}}},
  \bibinfo{author}{\bibfnamefont{Y.~P.} \bibnamefont{{Jing}}},
  \bibinfo{author}{\bibfnamefont{A.}~\bibnamefont{{Pasquali}}},
  \bibnamefont{and} \bibinfo{author}{\bibfnamefont{H.~J.} \bibnamefont{{Mo}}},
  \bibinfo{journal}{\apjl} \textbf{\bibinfo{volume}{662}}, \bibinfo{pages}{L71}
  (\bibinfo{year}{2007}), \eprint{0704.0674}.

\bibitem[{\citenamefont{Knebe et~al.}(2008)\citenamefont{Knebe, Draganova,
  Power, Yepes, Hoffman, Gottl{\"o}ber, and Gibson}}]{knebe08}
\bibinfo{author}{\bibfnamefont{A.}~\bibnamefont{Knebe}},
  \bibinfo{author}{\bibfnamefont{N.}~\bibnamefont{Draganova}},
  \bibinfo{author}{\bibfnamefont{C.}~\bibnamefont{Power}},
  \bibinfo{author}{\bibfnamefont{G.}~\bibnamefont{Yepes}},
  \bibinfo{author}{\bibfnamefont{Y.}~\bibnamefont{Hoffman}},
  \bibinfo{author}{\bibfnamefont{S.}~\bibnamefont{Gottl{\"o}ber}},
  \bibnamefont{and} \bibinfo{author}{\bibfnamefont{B.~K.}
  \bibnamefont{Gibson}}, \bibinfo{journal}{Monthly Notices RAS Letters}
  \textbf{\bibinfo{volume}{386}}, \bibinfo{pages}{L52} (\bibinfo{year}{2008}),
  \urlprefix\url{http://adsabs.harvard.edu/cgi-bin/nph-data_query?bibcode=2008%
MNRAS.386L..52K&link_type=ABSTRACT}.

\bibitem[{\citenamefont{Pereira et~al.}(2008)\citenamefont{Pereira, Bryan, and
  Gill}}]{pereira08}
\bibinfo{author}{\bibfnamefont{M.~J.} \bibnamefont{Pereira}},
  \bibinfo{author}{\bibfnamefont{G.~L.} \bibnamefont{Bryan}}, \bibnamefont{and}
  \bibinfo{author}{\bibfnamefont{S.~P.~D.} \bibnamefont{Gill}},
  \bibinfo{journal}{The Astrophysical Journal} \textbf{\bibinfo{volume}{672}},
  \bibinfo{pages}{825} (\bibinfo{year}{2008}), \bibinfo{note}{(c) 2008: The
  American Astronomical Society},
  \urlprefix\url{http://adsabs.harvard.edu/cgi-bin/nph-data_query?bibcode=2008%
ApJ...672..825P&link_type=ABSTRACT}.

\bibitem[{\citenamefont{Plionis et~al.}(2003)\citenamefont{Plionis, Benoist,
  Maurogordato, Ferrari, and Basilakos}}]{plionis03}
\bibinfo{author}{\bibfnamefont{M.}~\bibnamefont{Plionis}},
  \bibinfo{author}{\bibfnamefont{C.}~\bibnamefont{Benoist}},
  \bibinfo{author}{\bibfnamefont{S.}~\bibnamefont{Maurogordato}},
  \bibinfo{author}{\bibfnamefont{C.}~\bibnamefont{Ferrari}}, \bibnamefont{and}
  \bibinfo{author}{\bibfnamefont{S.}~\bibnamefont{Basilakos}},
  \bibinfo{journal}{The Astrophysical Journal} \textbf{\bibinfo{volume}{594}},
  \bibinfo{pages}{144} (\bibinfo{year}{2003}), \bibinfo{note}{(c) 2003: The
  American Astronomical Society},
  \urlprefix\url{http://adsabs.harvard.edu/cgi-bin/nph-data_query?bibcode=2003%
ApJ...594..144P&link_type=ABSTRACT}.

\bibitem[{\citenamefont{{Agustsson} and {Brainerd}}(2007)}]{agustsson07}
\bibinfo{author}{\bibfnamefont{I.}~\bibnamefont{{Agustsson}}} \bibnamefont{and}
  \bibinfo{author}{\bibfnamefont{T.~G.} \bibnamefont{{Brainerd}}},
  \bibinfo{journal}{ArXiv e-prints}  (\bibinfo{year}{2007}),
  \eprint{0704.3441}.

\bibitem[{\citenamefont{{Faltenbacher}
  et~al.}(2008)\citenamefont{{Faltenbacher}, {Jing}, {Li}, {Mao}, {Mo},
  {Pasquali}, and {van den Bosch}}}]{faltenbacher08}
\bibinfo{author}{\bibfnamefont{A.}~\bibnamefont{{Faltenbacher}}},
  \bibinfo{author}{\bibfnamefont{Y.~P.} \bibnamefont{{Jing}}},
  \bibinfo{author}{\bibfnamefont{C.}~\bibnamefont{{Li}}},
  \bibinfo{author}{\bibfnamefont{S.}~\bibnamefont{{Mao}}},
  \bibinfo{author}{\bibfnamefont{H.~J.} \bibnamefont{{Mo}}},
  \bibinfo{author}{\bibfnamefont{A.}~\bibnamefont{{Pasquali}}},
  \bibnamefont{and} \bibinfo{author}{\bibfnamefont{F.~C.} \bibnamefont{{van den
  Bosch}}}, \bibinfo{journal}{\apj} \textbf{\bibinfo{volume}{675}},
  \bibinfo{pages}{146} (\bibinfo{year}{2008}), \eprint{0706.0262}.

\bibitem[{\citenamefont{{Faltenbacher}
  et~al.}(2009)\citenamefont{{Faltenbacher}, {Li}, {White}, {Jing},
  {Shu-DeMao}, and {Wang}}}]{faltenbacher09}
\bibinfo{author}{\bibfnamefont{A.}~\bibnamefont{{Faltenbacher}}},
  \bibinfo{author}{\bibfnamefont{C.}~\bibnamefont{{Li}}},
  \bibinfo{author}{\bibfnamefont{S.~D.~M.} \bibnamefont{{White}}},
  \bibinfo{author}{\bibfnamefont{Y.-P.} \bibnamefont{{Jing}}},
  \bibinfo{author}{\bibnamefont{{Shu-DeMao}}}, \bibnamefont{and}
  \bibinfo{author}{\bibfnamefont{J.}~\bibnamefont{{Wang}}},
  \bibinfo{journal}{Research in Astronomy and Astrophysics}
  \textbf{\bibinfo{volume}{9}}, \bibinfo{pages}{41} (\bibinfo{year}{2009}),
  \eprint{0811.1995}.

\bibitem[{\citenamefont{{Siverd} et~al.}(2009)\citenamefont{{Siverd}, {Ryden},
  and {Gaudi}}}]{siverd09}
\bibinfo{author}{\bibfnamefont{R.~J.} \bibnamefont{{Siverd}}},
  \bibinfo{author}{\bibfnamefont{B.~S.} \bibnamefont{{Ryden}}},
  \bibnamefont{and} \bibinfo{author}{\bibfnamefont{B.~S.}
  \bibnamefont{{Gaudi}}}, \bibinfo{journal}{ArXiv e-prints}
  (\bibinfo{year}{2009}), \eprint{0903.2264}.

\bibitem[{\citenamefont{{Bridle} and {Abdalla}}(2007)}]{bridlea07}
\bibinfo{author}{\bibfnamefont{S.}~\bibnamefont{{Bridle}}} \bibnamefont{and}
  \bibinfo{author}{\bibfnamefont{F.~B.} \bibnamefont{{Abdalla}}},
  \bibinfo{journal}{\apjl} \textbf{\bibinfo{volume}{655}}, \bibinfo{pages}{L1}
  (\bibinfo{year}{2007}), \eprint{arXiv:astro-ph/0608002}.

\bibitem[{\citenamefont{{Brown} et~al.}(2002)\citenamefont{{Brown}, {Taylor},
  {Hambly}, and {Dye}}}]{brownthd02}
\bibinfo{author}{\bibfnamefont{M.~L.} \bibnamefont{{Brown}}},
  \bibinfo{author}{\bibfnamefont{A.~N.} \bibnamefont{{Taylor}}},
  \bibinfo{author}{\bibfnamefont{N.~C.} \bibnamefont{{Hambly}}},
  \bibnamefont{and} \bibinfo{author}{\bibfnamefont{S.}~\bibnamefont{{Dye}}},
  \bibinfo{journal}{\mnras} \textbf{\bibinfo{volume}{333}},
  \bibinfo{pages}{501} (\bibinfo{year}{2002}), \eprint{arXiv:astro-ph/0009499}.

\bibitem[{\citenamefont{Heymans et~al.}(2004)\citenamefont{Heymans, Brown,
  Heavens, Meisenheimer, Taylor, and Wolf}}]{heymans04}
\bibinfo{author}{\bibfnamefont{C.}~\bibnamefont{Heymans}},
  \bibinfo{author}{\bibfnamefont{M.}~\bibnamefont{Brown}},
  \bibinfo{author}{\bibfnamefont{A.}~\bibnamefont{Heavens}},
  \bibinfo{author}{\bibfnamefont{K.}~\bibnamefont{Meisenheimer}},
  \bibinfo{author}{\bibfnamefont{A.}~\bibnamefont{Taylor}}, \bibnamefont{and}
  \bibinfo{author}{\bibfnamefont{C.}~\bibnamefont{Wolf}},
  \bibinfo{journal}{Monthly Notices RAS} \textbf{\bibinfo{volume}{347}},
  \bibinfo{pages}{895} (\bibinfo{year}{2004}), \bibinfo{note}{(c) 2004 RAS},
  \urlprefix\url{http://adsabs.harvard.edu/cgi-bin/nph-data_query?bibcode=2004%
MNRAS.347..895H&link_type=ABSTRACT}.

\bibitem[{\citenamefont{Mandelbaum et~al.}(2006)\citenamefont{Mandelbaum,
  Hirata, Ishak, Seljak, and Brinkmann}}]{mandelbaum06}
\bibinfo{author}{\bibfnamefont{R.}~\bibnamefont{Mandelbaum}},
  \bibinfo{author}{\bibfnamefont{C.~M.} \bibnamefont{Hirata}},
  \bibinfo{author}{\bibfnamefont{M.}~\bibnamefont{Ishak}},
  \bibinfo{author}{\bibfnamefont{U.}~\bibnamefont{Seljak}}, \bibnamefont{and}
  \bibinfo{author}{\bibfnamefont{J.}~\bibnamefont{Brinkmann}},
  \bibinfo{journal}{Monthly Notices RAS} \textbf{\bibinfo{volume}{367}},
  \bibinfo{pages}{611} (\bibinfo{year}{2006}),
  \urlprefix\url{http://adsabs.harvard.edu/cgi-bin/nph-data_query?bibcode=2006%
MNRAS.367..611M&link_type=ABSTRACT}.

\bibitem[{\citenamefont{{Lee} and {Pen}}(2007)}]{leep07}
\bibinfo{author}{\bibfnamefont{J.}~\bibnamefont{{Lee}}} \bibnamefont{and}
  \bibinfo{author}{\bibfnamefont{U.-L.} \bibnamefont{{Pen}}},
  \bibinfo{journal}{\apjl} \textbf{\bibinfo{volume}{670}}, \bibinfo{pages}{L1}
  (\bibinfo{year}{2007}), \eprint{0707.3232}.

\bibitem[{\citenamefont{Hirata et~al.}(2007)\citenamefont{Hirata, Mandelbaum,
  Ishak, Seljak, Nichol, Pimbblet, Ross, and Wake}}]{hirata07}
\bibinfo{author}{\bibfnamefont{C.~M.} \bibnamefont{Hirata}},
  \bibinfo{author}{\bibfnamefont{R.}~\bibnamefont{Mandelbaum}},
  \bibinfo{author}{\bibfnamefont{M.}~\bibnamefont{Ishak}},
  \bibinfo{author}{\bibfnamefont{U.}~\bibnamefont{Seljak}},
  \bibinfo{author}{\bibfnamefont{R.}~\bibnamefont{Nichol}},
  \bibinfo{author}{\bibfnamefont{K.~A.} \bibnamefont{Pimbblet}},
  \bibinfo{author}{\bibfnamefont{N.~P.} \bibnamefont{Ross}}, \bibnamefont{and}
  \bibinfo{author}{\bibfnamefont{D.}~\bibnamefont{Wake}},
  \bibinfo{journal}{Monthly Notices RAS} \textbf{\bibinfo{volume}{381}},
  \bibinfo{pages}{1197} (\bibinfo{year}{2007}),
  \urlprefix\url{http://adsabs.harvard.edu/cgi-bin/nph-data_query?bibcode=2007%
MNRAS.381.1197H&link_type=ABSTRACT}.

\bibitem[{\citenamefont{{Kitching} et~al.}(2008)\citenamefont{{Kitching},
  {Amara}, {Abdalla}, {Joachimi}, and {Refregier}}}]{kitching09}
\bibinfo{author}{\bibfnamefont{T.~D.} \bibnamefont{{Kitching}}},
  \bibinfo{author}{\bibfnamefont{A.}~\bibnamefont{{Amara}}},
  \bibinfo{author}{\bibfnamefont{F.~B.} \bibnamefont{{Abdalla}}},
  \bibinfo{author}{\bibfnamefont{B.}~\bibnamefont{{Joachimi}}},
  \bibnamefont{and}
  \bibinfo{author}{\bibfnamefont{A.}~\bibnamefont{{Refregier}}},
  \bibinfo{journal}{ArXiv e-prints}  (\bibinfo{year}{2008}),
  \eprint{0812.1966}.

\bibitem[{\citenamefont{{King} and {Schneider}}(2003)}]{king02}
\bibinfo{author}{\bibfnamefont{L.~J.} \bibnamefont{{King}}} \bibnamefont{and}
  \bibinfo{author}{\bibfnamefont{P.}~\bibnamefont{{Schneider}}},
  \bibinfo{journal}{\aap} \textbf{\bibinfo{volume}{398}}, \bibinfo{pages}{23}
  (\bibinfo{year}{2003}), \eprint{arXiv:astro-ph/0209474}.

\bibitem[{\citenamefont{{Heymans} and {Heavens}}(2003)}]{heymans03}
\bibinfo{author}{\bibfnamefont{C.}~\bibnamefont{{Heymans}}} \bibnamefont{and}
  \bibinfo{author}{\bibfnamefont{A.}~\bibnamefont{{Heavens}}},
  \bibinfo{journal}{\mnras} \textbf{\bibinfo{volume}{339}},
  \bibinfo{pages}{711} (\bibinfo{year}{2003}), \eprint{arXiv:astro-ph/0208220}.

\bibitem[{\citenamefont{{Takada} and {White}}(2004)}]{takada04}
\bibinfo{author}{\bibfnamefont{M.}~\bibnamefont{{Takada}}} \bibnamefont{and}
  \bibinfo{author}{\bibfnamefont{M.}~\bibnamefont{{White}}},
  \bibinfo{journal}{\apjl} \textbf{\bibinfo{volume}{601}}, \bibinfo{pages}{L1}
  (\bibinfo{year}{2004}), \eprint{arXiv:astro-ph/0311104}.

\bibitem[{\citenamefont{{Joachimi} and {Schneider}}(2008)}]{joachimi08}
\bibinfo{author}{\bibfnamefont{B.}~\bibnamefont{{Joachimi}}} \bibnamefont{and}
  \bibinfo{author}{\bibfnamefont{P.}~\bibnamefont{{Schneider}}},
  \bibinfo{journal}{\aap} \textbf{\bibinfo{volume}{488}}, \bibinfo{pages}{829}
  (\bibinfo{year}{2008}), \eprint{0804.2292}.

\bibitem[{\citenamefont{Bridle and King}(2007)}]{bridle07}
\bibinfo{author}{\bibfnamefont{S.}~\bibnamefont{Bridle}} \bibnamefont{and}
  \bibinfo{author}{\bibfnamefont{L.}~\bibnamefont{King}}, \bibinfo{journal}{New
  J. Phys.} \textbf{\bibinfo{volume}{9}}, \bibinfo{pages}{444}
  (\bibinfo{year}{2007}).

\bibitem[{\citenamefont{Zhang}(2008)}]{zhang08}
\bibinfo{author}{\bibfnamefont{P.}~\bibnamefont{Zhang}},
  \bibinfo{journal}{arXiv} \textbf{\bibinfo{volume}{astro-ph}}
  (\bibinfo{year}{2008}), \bibinfo{note}{7 pages, 3 figures. To be submitted to
  ApJ}, \urlprefix\url{http://arxiv.org/abs/0811.0613v1}.

\bibitem[{\citenamefont{{Scherrer} and {Bertschinger}}(1991)}]{scherrerb91}
\bibinfo{author}{\bibfnamefont{R.~J.} \bibnamefont{{Scherrer}}}
  \bibnamefont{and}
  \bibinfo{author}{\bibfnamefont{E.}~\bibnamefont{{Bertschinger}}},
  \bibinfo{journal}{\apj} \textbf{\bibinfo{volume}{381}}, \bibinfo{pages}{349}
  (\bibinfo{year}{1991}).

\bibitem[{\citenamefont{{Scoccimarro} et~al.}(2001)\citenamefont{{Scoccimarro},
  {Sheth}, {Hui}, and {Jain}}}]{scoccimarroshj01}
\bibinfo{author}{\bibfnamefont{R.}~\bibnamefont{{Scoccimarro}}},
  \bibinfo{author}{\bibfnamefont{R.~K.} \bibnamefont{{Sheth}}},
  \bibinfo{author}{\bibfnamefont{L.}~\bibnamefont{{Hui}}}, \bibnamefont{and}
  \bibinfo{author}{\bibfnamefont{B.}~\bibnamefont{{Jain}}},
  \bibinfo{journal}{\apj} \textbf{\bibinfo{volume}{546}}, \bibinfo{pages}{20}
  (\bibinfo{year}{2001}), \eprint{arXiv:astro-ph/0006319}.

\bibitem[{\citenamefont{{Cooray} and {Sheth}}(2002)}]{coorays02}
\bibinfo{author}{\bibfnamefont{A.}~\bibnamefont{{Cooray}}} \bibnamefont{and}
  \bibinfo{author}{\bibfnamefont{R.}~\bibnamefont{{Sheth}}},
  \bibinfo{journal}{\physrep} \textbf{\bibinfo{volume}{372}},
  \bibinfo{pages}{1} (\bibinfo{year}{2002}), \eprint{arXiv:astro-ph/0206508}.

\bibitem[{\citenamefont{{Allgood} et~al.}(2006)\citenamefont{{Allgood},
  {Flores}, {Primack}, {Kravtsov}, {Wechsler}, {Faltenbacher}, and
  {Bullock}}}]{allgood06}
\bibinfo{author}{\bibfnamefont{B.}~\bibnamefont{{Allgood}}},
  \bibinfo{author}{\bibfnamefont{R.~A.} \bibnamefont{{Flores}}},
  \bibinfo{author}{\bibfnamefont{J.~R.} \bibnamefont{{Primack}}},
  \bibinfo{author}{\bibfnamefont{A.~V.} \bibnamefont{{Kravtsov}}},
  \bibinfo{author}{\bibfnamefont{R.~H.} \bibnamefont{{Wechsler}}},
  \bibinfo{author}{\bibfnamefont{A.}~\bibnamefont{{Faltenbacher}}},
  \bibnamefont{and} \bibinfo{author}{\bibfnamefont{J.~S.}
  \bibnamefont{{Bullock}}}, \bibinfo{journal}{\mnras}
  \textbf{\bibinfo{volume}{367}}, \bibinfo{pages}{1781} (\bibinfo{year}{2006}),
  \eprint{arXiv:astro-ph/0508497}.

\bibitem[{\citenamefont{{Navarro} et~al.}(1997)\citenamefont{{Navarro},
  {Frenk}, and {White}}}]{nfw}
\bibinfo{author}{\bibfnamefont{J.~F.} \bibnamefont{{Navarro}}},
  \bibinfo{author}{\bibfnamefont{C.~S.} \bibnamefont{{Frenk}}},
  \bibnamefont{and} \bibinfo{author}{\bibfnamefont{S.~D.~M.}
  \bibnamefont{{White}}}, \bibinfo{journal}{\apj}
  \textbf{\bibinfo{volume}{490}}, \bibinfo{pages}{493} (\bibinfo{year}{1997}),
  \eprint{astro-ph/9611107}.

\bibitem[{\citenamefont{{Catelan}
  et~al.}(2001{\natexlab{b}})\citenamefont{{Catelan}, {Kamionkowski}, and
  {Blandford}}}]{catelan01}
\bibinfo{author}{\bibfnamefont{P.}~\bibnamefont{{Catelan}}},
  \bibinfo{author}{\bibfnamefont{M.}~\bibnamefont{{Kamionkowski}}},
  \bibnamefont{and} \bibinfo{author}{\bibfnamefont{R.~D.}
  \bibnamefont{{Blandford}}}, \bibinfo{journal}{\mnras}
  \textbf{\bibinfo{volume}{320}}, \bibinfo{pages}{L7}
  (\bibinfo{year}{2001}{\natexlab{b}}), \eprint{arXiv:astro-ph/0005470}.

\bibitem[{\citenamefont{Cooray and Sheth}(2002)}]{cooray02}
\bibinfo{author}{\bibfnamefont{A.}~\bibnamefont{Cooray}} \bibnamefont{and}
  \bibinfo{author}{\bibfnamefont{R.}~\bibnamefont{Sheth}},
  \bibinfo{journal}{Physics Reports} \textbf{\bibinfo{volume}{372}},
  \bibinfo{pages}{1} (\bibinfo{year}{2002}),
  \urlprefix\url{http://adsabs.harvard.edu/cgi-bin/nph-data_query?bibcode=2002%
PhR...372....1C&link_type=ABSTRACT}.

\bibitem[{\citenamefont{{Song} and {Knox}}(2004)}]{song04}
\bibinfo{author}{\bibfnamefont{Y.-S.} \bibnamefont{{Song}}} \bibnamefont{and}
  \bibinfo{author}{\bibfnamefont{L.}~\bibnamefont{{Knox}}},
  \bibinfo{journal}{\prd} \textbf{\bibinfo{volume}{70}},
  \bibinfo{pages}{063510} (\bibinfo{year}{2004}),
  \eprint{arXiv:astro-ph/0312175}.

\bibitem[{\citenamefont{{Bunn} et~al.}(2003)\citenamefont{{Bunn},
  {Zaldarriaga}, {Tegmark}, and {de Oliveira-Costa}}}]{bunn03}
\bibinfo{author}{\bibfnamefont{E.~F.} \bibnamefont{{Bunn}}},
  \bibinfo{author}{\bibfnamefont{M.}~\bibnamefont{{Zaldarriaga}}},
  \bibinfo{author}{\bibfnamefont{M.}~\bibnamefont{{Tegmark}}},
  \bibnamefont{and} \bibinfo{author}{\bibfnamefont{A.}~\bibnamefont{{de
  Oliveira-Costa}}}, \bibinfo{journal}{\prd} \textbf{\bibinfo{volume}{67}},
  \bibinfo{pages}{023501} (\bibinfo{year}{2003}),
  \eprint{arXiv:astro-ph/0207338}.

\bibitem[{\citenamefont{{Smail} et~al.}(1994)\citenamefont{{Smail}, {Ellis},
  and {Fitchett}}}]{smailef94}
\bibinfo{author}{\bibfnamefont{I.}~\bibnamefont{{Smail}}},
  \bibinfo{author}{\bibfnamefont{R.~S.} \bibnamefont{{Ellis}}},
  \bibnamefont{and} \bibinfo{author}{\bibfnamefont{M.~J.}
  \bibnamefont{{Fitchett}}}, \bibinfo{journal}{\mnras}
  \textbf{\bibinfo{volume}{270}}, \bibinfo{pages}{245} (\bibinfo{year}{1994}).

\bibitem[{\citenamefont{{Benjamin} et~al.}(2007)\citenamefont{{Benjamin},
  {Heymans}, {Semboloni}, {van Waerbeke}, {Hoekstra}, {Erben}, {Gladders},
  {Hetterscheidt}, {Mellier}, and {Yee}}}]{benjamin07}
\bibinfo{author}{\bibfnamefont{J.}~\bibnamefont{{Benjamin}}},
  \bibinfo{author}{\bibfnamefont{C.}~\bibnamefont{{Heymans}}},
  \bibinfo{author}{\bibfnamefont{E.}~\bibnamefont{{Semboloni}}},
  \bibinfo{author}{\bibfnamefont{L.}~\bibnamefont{{van Waerbeke}}},
  \bibinfo{author}{\bibfnamefont{H.}~\bibnamefont{{Hoekstra}}},
  \bibinfo{author}{\bibfnamefont{T.}~\bibnamefont{{Erben}}},
  \bibinfo{author}{\bibfnamefont{M.~D.} \bibnamefont{{Gladders}}},
  \bibinfo{author}{\bibfnamefont{M.}~\bibnamefont{{Hetterscheidt}}},
  \bibinfo{author}{\bibfnamefont{Y.}~\bibnamefont{{Mellier}}},
  \bibnamefont{and} \bibinfo{author}{\bibfnamefont{H.~K.~C.}
  \bibnamefont{{Yee}}}, \bibinfo{journal}{\mnras}
  \textbf{\bibinfo{volume}{381}}, \bibinfo{pages}{702} (\bibinfo{year}{2007}),
  \eprint{arXiv:astro-ph/0703570}.

\bibitem[{\citenamefont{{Huterer} et~al.}(2006)\citenamefont{{Huterer},
  {Takada}, {Bernstein}, and {Jain}}}]{huterertbj06}
\bibinfo{author}{\bibfnamefont{D.}~\bibnamefont{{Huterer}}},
  \bibinfo{author}{\bibfnamefont{M.}~\bibnamefont{{Takada}}},
  \bibinfo{author}{\bibfnamefont{G.}~\bibnamefont{{Bernstein}}},
  \bibnamefont{and} \bibinfo{author}{\bibfnamefont{B.}~\bibnamefont{{Jain}}},
  \bibinfo{journal}{\mnras} \textbf{\bibinfo{volume}{366}},
  \bibinfo{pages}{101} (\bibinfo{year}{2006}), \eprint{astro-ph/0506030}.

\bibitem[{\citenamefont{{Schneider} et~al.}(2002)\citenamefont{{Schneider},
  {van Waerbeke}, and {Mellier}}}]{schneider02}
\bibinfo{author}{\bibfnamefont{P.}~\bibnamefont{{Schneider}}},
  \bibinfo{author}{\bibfnamefont{L.}~\bibnamefont{{van Waerbeke}}},
  \bibnamefont{and}
  \bibinfo{author}{\bibfnamefont{Y.}~\bibnamefont{{Mellier}}},
  \bibinfo{journal}{\aap} \textbf{\bibinfo{volume}{389}}, \bibinfo{pages}{729}
  (\bibinfo{year}{2002}), \eprint{arXiv:astro-ph/0112441}.

\bibitem[{\citenamefont{Abramowitz and Stegun}(1964)}]{abromowitz+stegun}
\bibinfo{author}{\bibfnamefont{M.}~\bibnamefont{Abramowitz}} \bibnamefont{and}
  \bibinfo{author}{\bibfnamefont{I.~A.} \bibnamefont{Stegun}},
  \emph{\bibinfo{title}{Handbook of Mathematical Functions with Formulas,
  Graphs, and Mathematical Tables}} (\bibinfo{publisher}{Dover},
  \bibinfo{year}{1964}).

\end{thebibliography}
